\documentclass[12pt]{article}
\pdfoutput=1

\usepackage{amssymb}
\usepackage{amsmath}
\usepackage{bbold}
\usepackage{color}
\usepackage{colordvi}
\usepackage{colortbl}
\usepackage{fancybox}
\usepackage[footnotesize]{caption2}
\usepackage{graphicx}
\usepackage[center,footnotesize,hang]{subfigure}
\usepackage{bbm}
\usepackage{url}
\usepackage{multirow}
\usepackage{array}
\usepackage{arydshln}
\usepackage{pifont}
\usepackage{mathrsfs}
\usepackage{fancybox}
\usepackage{cite}
\usepackage[colorlinks]{hyperref}
\usepackage{enumitem}  
\usepackage{float} %
\newcommand{\PreserveBackslash}[1]{\let\temp=\\#1\let\\=\temp}
\newcolumntype{C}[1]{>{\PreserveBackslash\centering}p{#1}}
\newcolumntype{R}[1]{>{\PreserveBackslash\raggedleft}p{#1}}
\newcolumntype{L}[1]{>{\PreserveBackslash\raggedright}p{#1}}
\allowdisplaybreaks \allowdisplaybreaks[2]

\definecolor{lightred}{rgb}{1,0.4,0.4}

\begin{document}

\title{
\begin{flushright}
\hfill\mbox{\small USTC-ICTS-18-12} \\[5mm]
\begin{minipage}{0.2\linewidth}
\normalsize
\end{minipage}
\end{flushright}
{\Large \bf Tri-Direct CP in the Littlest Seesaw Playground
\\[2mm]}
}

\date{}

\author{
Gui-Jun~Ding$^{1}$\footnote{E-mail: {\tt
dinggj@ustc.edu.cn}},  \
Stephen~F.~King$^{2}$\footnote{E-mail: {\tt king@soton.ac.uk}}, \
Cai-Chang Li$^{1}$\footnote{E-mail: {\tt
lcc0915@mail.ustc.edu.cn}}  \
\\*[20pt]
\centerline{
\begin{minipage}{\linewidth}
\begin{center}
$^1${\it \small
Interdisciplinary Center for Theoretical Study and  Department of Modern Physics,\\
University of Science and Technology of China, Hefei, Anhui 230026, China}\\[2mm]
$^2${\it \small
Physics and Astronomy,
University of Southampton,
Southampton, SO17 1BJ, U.K.}\\
\end{center}
\end{minipage}}
\\[10mm]}
\maketitle
\thispagestyle{empty}

\begin{abstract}
\noindent
We discuss spontaneously broken CP symmetry in two right-handed neutrino models based on the idea
of having a {\it different residual flavour symmetry}, together with a {\it different residual CP symmetry},
associated with each of the two right-handed neutrinos. The charged lepton sector also has a {\it different residual flavour symmetry}.
In such a {\it tri-direct CP approach},
we show that the combination of the three residual flavour and two residual CP symmetries provides a new way of fixing the
parameters. To illustrate the approach, we revisit the Littlest Seesaw (LSS) model
based on $S_4$ and then propose new variants
which have not so far appeared in the literature, with different predictions for each variant. We analyse numerically the predictions of the new variants, and then propose an explicit model which can realise one of the successful benchmark points, based on the atmospheric
flavon vacuum alignment $(1, \omega^2 , \omega)$ and the solar flavon vacuum alignment
$(1, -7/2, -7/2 )$.
\end{abstract}
\newpage

\section{\label{sec:introduction}Introduction}
\indent

Following the discovery of neutrino mass and mixing, we are now firmly in the precision era of measurements. In the standard parametrisation of the lepton mixing matrix~\cite{Patrignani:2016xqp}, all the three lepton mixing angles $\theta_{12}$, $\theta_{13}$ and $\theta_{23}$ and the mass squared differences $\Delta m^2_{21}\equiv m^2_2-m^2_1$ and $\Delta m^2_{31}\equiv m^2_3-m^2_1$ has been precisely measured in a large number of neutrino oscillation experiments. At present the $3\sigma$ ranges of these mixing parameters are determined to be~\cite{Capozzi:2017ipn,deSalas:2017kay,Esteban:2016qun}
\begin{eqnarray}
\label{eq:3sigma_data}&&\hskip-0.1in 0.272\leq\sin^2\theta_{12}\leq0.346,\quad 0.01981\leq\sin^2\theta_{13}\leq0.02436,\quad
0.418\leq\sin^2\theta_{23}\leq0.613,\\
\nonumber&&\hskip-0,1in6.80\times10^{-5}\text{eV}^2\leq\Delta m^2_{21}\leq8.02\times10^{-5}\text{eV}^2,\quad
2.399\times10^{-3}\text{eV}^2\leq\Delta m^2_{31}\leq2.593\times10^{-3}\text{eV}^2\,,
\end{eqnarray}
where these results are as quoted in \cite{Esteban:2016qun}
for normal ordering (NO) neutrino mass spectrum, and similar results are obtained for inverted ordering (IO) spectrum except that the sign of $\Delta m^2_{31}$ is reversed.

Non-Abelian discrete finite family symmetry groups $G_f$ have been widely used to explain the lepton mixing angles as well as CP violating phases, see Refs.~\cite{Altarelli:2010gt,Ishimori:2010au,King:2013eh,King:2014nza,King:2015aea,King:2017guk} for reviews. One of the most successful and popular model independent approaches is to impose a discrete family symmetry $G_f$ together with a non-commuting CP symmetry $H_{CP}$. In the {\em semi-direct CP approach}, the $G_f\rtimes H_{CP}$ symmetry is subsequently spontaneously broken, leaving residual symmetries $G_{\nu}\rtimes H^{\nu}_{CP}$ in the neutrino sector and $G_{l}\rtimes H^{l}_{CP}$ in the charged lepton sector, leading to mixing angle and CP phase predictions. In the present paper we shall generalise the {\em semi-direct CP approach} to a {\em tri-direct CP approach}, based on the two right-handed neutrino seesaw mechanism, as we now discuss.

The most appealing possibility for the origin of neutrino mass
seems to be the seesaw mechanism which, in its original formulation, involves heavy right-handed Majorana neutrinos~\cite{Minkowski:1977sc,GellMann:1980vs,Yanagida:1979as,Glashow:1979nm,Mohapatra:1979ia,Schechter:1980gr}.
The most minimal version of the seesaw mechanism involves one \cite{King:1998jw} or two right-handed neutrinos~\cite{King:1999mb}. In order to reduce the number of free parameters still further to the smallest number possible, and hence increase predictivity, various
approaches to the two right-handed neutrino seesaw model have been suggested, such as postulating one~\cite{King:2002nf} or two~\cite{Frampton:2002qc} texture zeroes, however such two texture zero models are now phenomenologically excluded~\cite{Harigaya:2012bw} for the case of a normal neutrino mass hierarchy.

The minimal successful seesaw scheme with normal hierarchy is called the Littlest Seesaw (LSS) model~\cite{King:2013iva,King:2015dvf,King:2016yvg}. The LSS model corresponds to a two right-handed neutrino models with a particularly simple pattern of Dirac mass matrix elements in the basis where both the charged lepton mass matrix and the two-right-handed neutrino mass matrix are diagonal. The Dirac mass matrix involves only one texture zero, but the number of parameters is reduced dramatically since each column of this matrix is controlled by a single parameter. In practice this is achieved by introducing a Non-Abelian discrete family symmetry, which is spontaneously broken by flavon fields with particular vacuum alignments governed by remnant subgroups of the family symmetry.
This leads to a highly constrained model which is remarkably consistent with current data, but which can be tested in forthcoming neutrino experiments~\cite{Ballett:2016yod}. The LSS approach may also be incorporated into grand unified models~\cite{Bjorkeroth:2015ora,Bjorkeroth:2015uou,Bjorkeroth:2015tsa,Bjorkeroth:2017ybg}.

Originally the LSS model was formulated in the {\it indirect approach} based on a family symmetry to give the required vacuum alignments, but without any residual symmetry in the neutrino or charged lepton sector~\cite{King:2013iva}.
Later it was realised that it preserves a {\it different residual flavour symmetry for each flavon},
in the diagonal mass basis of two right-handed neutrinos, leading to a highly predictive set of possible alignments~\cite{King:2015dvf}. Most recently it was understood that the LSS model may arise
from a {\it semi-direct symmetry} approach corresponding to a
{\it different residual flavour symmetry for each charge sector}, where a particular residual flavour symmetry may be assumed in each of the neutrino and charged lepton sectors. To be precise, in the {\it semi-direct symmetry} approach, it was shown that there is an $SU$ subgroup of $S_4$ in the neutrino sector and the $T$ subgroup of $S_4$ in the charged lepton sector,
leading to a constrained form of TM1 mixing~\cite{King:2016yvg} in which the first column of the tri-bimaximal mixing matrix is preserved, but with the reactor angle and CP phases fixed by the same two parameters which fix the neutrino masses.

The LSS model is a general and predictive framework for explaining neutrino masses and lepton mixing, and it is not confined to TM1. For instance, the golden LSS is an another viable class of LSS models~\cite{Ding:2017hdv}, the flavor symmetry group is $A_5$ and it is spontaneously broken to different residual subgroups in the charged lepton, atmospheric neutrino and solar neutrino sectors. The golden LSS predicts the lepton mixing is of GR1 form where the first column of the golden ratio mixing matrix is preserved~\cite{Ding:2017hdv}. In both the original LSS and golden LSS models, it was always assumed that there is a high energy CP symmetry which is completely broken in each of the sectors, with no residual CP symmetry.

\begin{figure}[t!]
\centering
\begin{tabular}{c}
\includegraphics[width=0.50\linewidth]{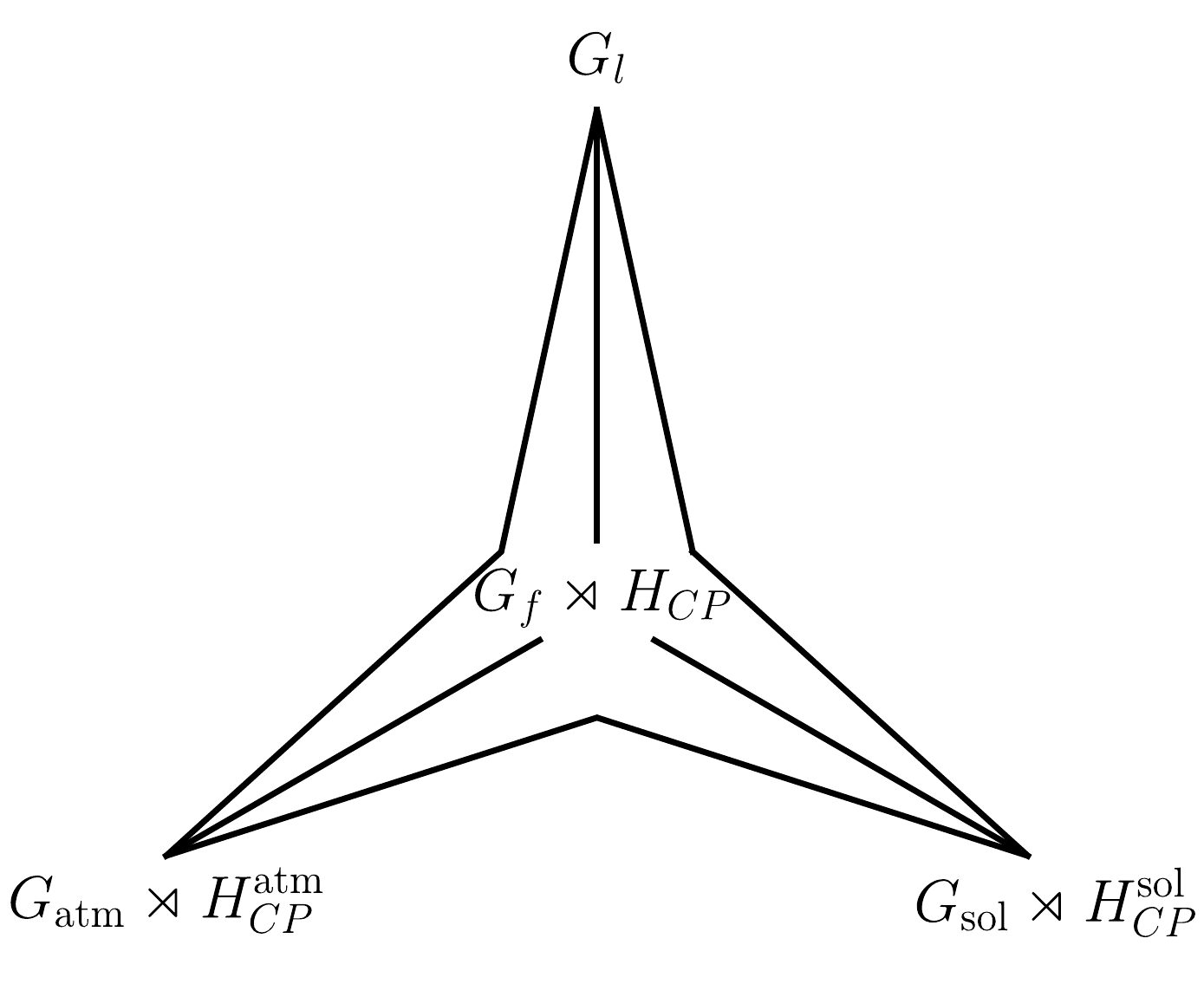}
\end{tabular}
\caption{\label{fig:benz} A sketch of the {\it tri-direct CP approach} for two right-handed neutrino models, where
the high energy family and CP symmetry $G_f\rtimes H_{CP}$
is spontaneously broken down to $G_{\text{atm}}\rtimes H_{CP}^{\text{atm}}$ in the sector of one of the right-handed neutrinos,
and $G_{\text{sol}}\rtimes H_{CP}^{\text{sol}}$ in the sector of the other right-handed neutrino, with
the charged lepton sector having a {\it different residual flavour symmetry} $G_{l}$.}
\end{figure}

In this paper we propose a new {\it tri-direct CP approach} for two right-handed neutrino models based on the idea
of spontaneously broken family and CP symmetry, leaving a {\it different residual flavour symmetry}, together with a {\it different residual CP symmetry},
in each of the two right-handed neutrino sectors. In other words, the high energy family and CP symmetry $G_f\rtimes H_{CP}$
is spontaneously broken down to $G_{\text{atm}}\rtimes H_{CP}^{\text{atm}}$ in the sector of one of the right-handed neutrinos,
and $G_{\text{sol}}\rtimes H_{CP}^{\text{sol}}$ in the sector of the other right-handed neutrino, with
the charged lepton sector having a {\it different residual flavour symmetry} $G_{l}$, as schematically illustrated in figure~\ref{fig:benz}. The tri-direct CP approach is a hybrid of the direct and indirect approaches. The common residual symmetry of the neutrino sector in the direct model is splitted into two branches: the residual symmetries associated with the atmospheric and solar neutrinos. In comparison with the indirect model, the alignments associated with each right-handed neutrino are enforced by residual symmetry. In such a {\it tri-direct CP approach} the combination of the three residual symmetries provides a new way of fixing the
parameters. To illustrate the approach, we revisit the Littlest Seesaw (LSS) model
based on $S_4$ and show that the {\it tri-direct CP approach} uniquely fixes some
parameters of the model.\footnote{This is similar to having separate residual symmetries for each right-handed neutrino arising from
$S_4$, as in~\cite{King:2015dvf}, but here we also impose separate residual CP symmetries.}
Following the {\it tri-direct CP approach}, we also propose new variants of the LSS model
which have not so far appeared in the literature, with different predictions for each variant. We analyse numerically the predictions of the new variants, and then propose an explicit model which can realise one of the successful benchmark points, based on the atmospheric
flavon vacuum alignment $(1, \omega^2 , \omega)$ and the solar flavon vacuum alignment
$(1, -7/2, -7/2 )$.

The layout of this paper is as follows. In section~\ref{sec:framework} we propose the tri-direct CP approach for two right-handed neutrino models.
In section~\ref{LSStri-direct} we apply the tri-direct CP approach to the Littlest Seesaw model and see that it reproduces the usual neutrino mass matrices arising from uniquely fixed vacuum alignments. In section~\ref{sec:new_LSS} we show how new variants of the Littlest Seesaw emerge from the tri-direct CP approach, and we perform a comprehensive numerical analysis of a selection of benchmark points within the LSS variants arising from $S_4$, in order to determine their viability and predictions. In section~\ref{sec:extension} the tri-direct CP approach is extended to three right-handed neutrino models. In section~\ref{sec:model} we propose an explicit model which can realise one of the successful benchmark points, based on the atmospheric
flavon vacuum alignment $(1, \omega^2 , \omega)$ and the solar flavon vacuum alignment $(1, -7/2, -7/2 )$. Section~\ref{sec:Conclusion} concludes the paper. The Appendix~\ref{sec:appendix_A} describes the diagonalization of a general subdiagonal neutrino mass matrix, note that the neutrino mass matrix predicted in the tri-direct CP approach can always be reduced to a subdiagonal one by performing a unitary transformation.

\section{\label{sec:framework} The tri-direct CP approach}

In a two right-handed neutrino model, the light neutrino masses are generated through the seesaw mechanism, and only two right-handed neutrinos are introduced,
denoted here as $N^c_{\mathrm{atm}}$ and $N^c_{\mathrm{sol}}$. In the right-handed neutrino diagonal basis, the most general Lagrangian can be written as
\begin{small}
\begin{equation}\label{eq:Lagrangian}
\mathcal{L}=-y_{l}L\phi_{l}E^{c}-y_{\mathrm{atm}}L\phi_{\mathrm{atm}}N^c_{\mathrm{atm}}-y_{\mathrm{sol}}L\phi_{\mathrm{sol}}N^c_{\mathrm{sol}}
-\frac{1}{2}x_{\mathrm{atm}}\xi_{\mathrm{atm}}N^c_{\mathrm{atm}}N^c_{\mathrm{atm}}-\frac{1}{2}x_{\mathrm{sol}}\xi_{\mathrm{sol}}N^{c}_{\mathrm{sol}}N^c_{\mathrm{sol}}
+\text{h.c.}\,,
\end{equation}
\end{small}
where we use a two-component notation for the fermion fields to keep the formula compact. The fields $L$ and $E^c\equiv(e^{c}, \mu^{c}, \tau^{c})^T$ stand for the left-handed lepton doublets and the right-handed charged leptons respectively, the flavons $\phi_{l}$, $\phi_{\rm sol}$ and $\phi_{\rm atm}$ can be either Higgs fields or combinations of the electroweak Higgs doublet together with flavons, and the Majoron flavons $\xi_{\text{atm}}$ and $\xi_{\text{sol}}$ are standard model singlets.

In order to predict both neutrino masses and lepton mixing parameters, a non-abelian discrete flavor symmetry $G_f$ and generalized CP symmetry $H_{CP}$ are imposed on the model. Both flavor symmetry and the CP symmetry act on the flavor space in a non-trivial way. The flavor symmetry $G_f$
and CP symmetry $H_{CP}$ should be compatible with each other in the theory, and consequently the following consistency condition has to be satisfied~\cite{Feruglio:2012cw,Holthausen:2012dk,Chen:2014tpa}
\begin{equation}
\label{eq:consistency_condition}X_{\mathbf{r}}\rho^{*}_{\mathbf{r}}(g)X^{\dagger}_{\mathbf{r}}=\rho_{\mathbf{r}}(g^{\prime}),\quad
g,g^{\prime}\in G_{f}\,,
\end{equation}
where $\rho_{\mathbf{r}}(g)$ is the representation matrix of the element $g$ in the irreducible representation $\mathbf{r}$ of $G_{f}$, and $X_{\mathbf{r}}$ is the generalized CP transformation matrix of $H_{CP}$. The elements $g$ and $g^{\prime}$ in Eq.~\eqref{eq:consistency_condition} are in general different. Hence the mathematical structure of the full symmetry at high energy scale is in general a semi-direct product of flavor symmetry $G_f$ and generalized CP symmetry $H_{CP}$~\cite{Feruglio:2012cw}.
Moreover, it has been shown that physical CP transformations always have to be class-inverting automorphisms of $G_f$~\cite{Chen:2014tpa}, namely for any $g\in G_{f}$, there should always exists an element $h$ such that $g'=hg^{-1}h^{-1}$. We assign the three generations of left-handed leptons $L$ to a faithful irreducible three-dimensional representation of $G_{f}$,
the flavons $\phi_{l}$, $\phi_{\text{atm}}$ and $\phi_{\text{sol}}$ transform as triplets under the flavor symmetry $G_{f}$ while
$N^c_{\mathrm{atm}}$ and $N^c_{\mathrm{sol}}$ are singlets of $G_{f}$. The flavons $\xi_{\textrm{atm}}$ and $\xi_{\textrm{sol}}$ are also singlets under $G_{f}$, nevertheless they could transform differently under the shaping symmetry. In the present work, we assume the parent symmetry $G_f\rtimes H_{CP}$ is broken down to $G_{l}$, $G_{\text{atm}}\rtimes H^{\text{atm}}_{CP}$ and $G_{\text{sol}}\rtimes H^{\text{sol}}_{CP}$ in the charged lepton, atmospheric neutrino and solar neutrino sectors, respectively. This paradigm is schematically illustrated in figure~\ref{fig:benz}, and we call it as tri-direct CP approach. Notice that the tri-direct CP approach has three branches of residual symmetries, and it is a generalization of the so called direct approach in which the flavor symmetry is broken to two distinct abelian subgroups $G_{l}$ and $G_{\nu}$ in the changed lepton and neutrino sectors respectively. Here we require that the residual flavor symmetry $G_{l}$ is capable of distinguishing the three generations, i.e., the order of $G_{l}$ can not be smaller than $3$. The invariance of the charged lepton mass term under the action of $G_{l}$ leads to
\begin{equation}
g^{\dagger}_{l}m^{\dagger}_{l}m_{l}g_{l}=m^{\dagger}_{l}m_{l}, \quad g_{l}\in G_{l}\,,
\end{equation}
which implies
\begin{equation}
\label{eq:commutable}[g_{l}, m^{\dagger}_{l}m_{l}]=0\,,
\end{equation}
where for simplicity we have used the same notations for the abstract elements of $G_f$ and the representation matrices in the triplet representation to which the lepton doublets $L$ are assigned. In addition,  the charged lepton mass matrix $m_l$ defined in the right-left basis $l^{c}m_ll$. The representation matrix $g_{l}$ can be diagonalized by a unitary transformation $U_{l}$, i.e.,
\begin{equation}
U^{\dagger}_{l}g_{l}U_{l}=\text{diag}(e^{i\alpha_e},e^{i\alpha_\mu}, e^{i\alpha_\tau})\,,
\end{equation}
where $e^{i\alpha_{e,\mu,\tau}}$ are all roots of unity since $g_{l}$ is an element of the discrete flavor symmetry group $G_{f}$ and it is of finite order, and $U_l$ is determined up to exchange of its column and possible overall phases of the
single columns. From Eq.~\eqref{eq:commutable}, it follows that the hermitian combination $m^{\dagger}_{l}m_{l}$ is diagonalized by $U_{l}$ as well.

As regards the atmospheric and solar neutrino sectors, the residual CP symmetry should be compatible with the residual flavor symmetry. As a consequence, the following constrained consistency conditions have to be fulfilled,
\begin{subequations}
\begin{eqnarray}
\label{eq:nu_atm_consistency}&&X^{\text{atm}}_{\mathbf{r}}\rho^{*}_{\mathbf{r}}(g^{\text{atm}}_{i})(X^{\text{atm}}_{\mathbf{r}})^{-1}
=\rho_{\mathbf{r}}(g^{\text{atm}}_{j}),\qquad g^{\text{atm}}_{i},g^{\text{atm}}_{j}\in G_{\text{atm}}\,,\\
\label{eq:nu_sol_consistency}&&X^{\text{sol}}_{\mathbf{r}}\rho^{*}_{\mathbf{r}}(g^{\text{sol}}_{i})(X^{\text{sol}}_{\mathbf{r}})^{-1}
=\rho_{\mathbf{r}}(g^{\text{sol}}_{j}),\qquad g^{\text{sol}}_{i},g^{\text{sol}}_{j}\in G_{\text{sol}}\,.
\end{eqnarray}
\end{subequations}
This implies that the mathematical structure of the subgroup comprising the residual flavor and CP symmetries is in general a semi-direct product. The semi-direct product structure will reduce to a direct product if $g^{\text{atm}}_{i}=g^{\text{atm}}_{j}$ and $g^{\text{sol}}_{i}=g^{\text{sol}}_{j}$. In particular, we note that the reduction of the semidirect product structure to direct product always holds true for a generic residual $Z_2$ flavor symmetry. For given residual flavor symmetries $G_{\text{atm}}$ and $G_{\text{sol}}$, the residual CP symmetries $H^{\text{atm}}_{CP}$ and $H^{\text{sol}}_{CP}$ can be easily obtained by solving the constraints in Eqs.~\eqref{eq:nu_atm_consistency} and \eqref{eq:nu_sol_consistency}. The requirement that the subgroup $G_{\text{atm}}\rtimes H^{\text{atm}}_{CP}$ is conserved entails that the vacuum expectation value
(VEV) of $\phi_{\text{atm}}$ should be invariant under the symmetry $G_{\text{atm}}\rtimes H^{\text{atm}}_{CP}$, i.e.
\begin{equation}
\begin{aligned}
&g_{\text{atm}}\langle\phi_{\text{atm}}\rangle=\langle\phi_{\text{atm}}\rangle,\qquad g_{\text{atm}}\in G_{\text{atm}}\,,\\
&X_{\text{atm}}\langle\phi_{\text{atm}}\rangle^{*}=\langle\phi_{\text{atm}}\rangle,\quad X_{\text{atm}}\in H^{\text{atm}}_{CP}\,,
\end{aligned}
\end{equation}
which allow us to fix the alignment of $\langle\phi_{\text{atm}}\rangle$. In the same fashion, for the symmetry $G_{\text{sol}}\rtimes H^{\text{sol}}_{CP}$ to hold, the VEV $\langle\phi_{\text{sol}}\rangle$ has to be invariant under $G_{\text{sol}}\rtimes H^{\text{sol}}_{CP}$,
\begin{equation}
\begin{aligned}
&g_{\text{sol}}\langle\phi_{\text{sol}}\rangle=\langle\phi_{\text{sol}}\rangle,\qquad g_{\text{sol}}\in G_{\text{sol}}\,,\\
&X_{\text{sol}}\langle\phi_{\text{sol}}\rangle^{*}=\langle\phi_{\text{sol}}\rangle,\quad X_{\text{sol}}\in H^{\text{sol}}_{CP}\,.
\end{aligned}
\end{equation}
After the electroweak and flavor symmetry breaking, the flavon VEVs $\langle\phi_{l}\rangle$, $\langle\phi_{\text{atm}}\rangle$,  $\langle\phi_{\text{sol}}\rangle$, $\langle\xi_{\text{atm}}\rangle$ and $\langle\xi_{\text{sol}}\rangle$ are non-vanishing. Then we can read out the neutrino Dirac mass matrix and the heavy Majorana mass matrix of $N^c_{\mathrm{atm}}$ and $N^c_{\mathrm{sol}}$,
\begin{equation}
m_{D}=\begin{pmatrix}
y_{\text{atm}}\langle\phi_{\text{atm}}\rangle,  ~&~ y_{\text{sol}}\langle\phi_{\text{sol}}\rangle
\end{pmatrix},\qquad m_{N}=\begin{pmatrix}
x_{\textrm{atm}}\langle\xi_{\text{atm}}\rangle ~&~ 0 \\
0  ~& ~ x_{\textrm{sol}}\langle\xi_{\text{sol}}\rangle
\end{pmatrix}\,,
\end{equation}
where we omit the relevant Clebsch-Gordan coefficients which appear in both contractions $y_{\mathrm{atm}}L\phi_{\mathrm{atm}}N^c_{\mathrm{atm}}$ and $y_{\mathrm{sol}}L\phi_{\mathrm{sol}}N^c_{\mathrm{sol}}$ in order to form invariants under $G_f$. Using the seesaw formula, we can obtain the low energy effective light neutrino
mass matrix
\begin{equation}
m_{\nu}=-\frac{y^2_{\text{atm}}}{x_{\text{atm}}}\frac{\langle\phi_{\text{atm}}\rangle\langle\phi_{\text{atm}}\rangle^{T}}{\langle\xi_{\text{atm}}\rangle}
-\frac{y^2_{\text{sol}}}{x_{\text{sol}}}\frac{\langle\phi_{\text{sol}}\rangle\langle\phi_{\text{sol}}\rangle^{T}}{\langle\xi_{\text{sol}}\rangle}\,.
\end{equation}
Since two right-handed neutrinos are introduced in this approach, the lightest neutrino must be massless. Indeed we find the light neutrino mass  matrix satisfy
\begin{equation}
m_{\nu}\left[\langle\phi_{\text{atm}}\rangle\times\langle\phi_{\text{sol}}\rangle\right]=(0, 0, 0)^{T}\,,
\end{equation}
where $\langle\phi_{\text{atm}}\rangle\times\langle\phi_{\text{sol}}\rangle$ denotes the cross product of $\langle\phi_{\text{atm}}\rangle$ and $\langle\phi_{\text{sol}}\rangle$. This means that the column vector $\langle\phi_{\text{atm}}\rangle\times\langle\phi_{\text{sol}}\rangle$ is an eigenvector of $m_{\nu}$ with zero eigenvalue, depending on the modulus of each entry, it can be either the first column or the third column of the neutrino unitary transformation matrix $U_{\nu}$ which diagonalizes $m_{\nu}$. Accordingly the neutrino mass spectrum can be normal order with $m_1=0$ or inverted ordering with $m_3=0$. The other two remaining columns of $U_{\nu}$ are orthogonal to $\langle\phi_{\text{atm}}\rangle\times\langle\phi_{\text{sol}}\rangle$, and their exact forms can be determined for given residual symmetry in a model, as explicitly shown in the following two sections. We can thus determine the lepton mixing matrix $U_{PMNS}=U^{\dagger}_lU_{\nu}$. Furthermore, we notice that the two columns of the Dirac mass matrix $m_{D}$ would be exchanged if the roles of $G_{\text{atm}}\rtimes H^{\text{atm}}_{CP}$ and $G_{\text{sol}}\rtimes H^{\text{sol}}_{CP}$ are switched. Thus the same neutrino mass matrix would be obtained if one interchanges $y_{\text{atm}}$ with $y_{\text{sol}}$ and $x_{\text{atm}}$ with $x_{\text{sol}}$. The tri-direct CP approach provides new opportunity for model building, it allows us to construct quite predictive neutrino mass models as simple as the LSS model.

\section{\label{LSStri-direct}Littlest Seesaw in the tri-direct CP approach }

In this section, we shall show that the original Littlest seesaw model~\cite{King:2015dvf,King:2016yvg} can be reproduced from the above tri-direct approach. The flavor symmetry group is chosen to be $S_4$ which has been extensively studied in the literature, see~\cite{Ding:2009iy,Ding:2013hpa} for the group theory of $S_4$. In the present work, we shall adopt the conventions of~\cite{Ding:2013hpa}, and $S_4$ can be generated by three generators $S$, $T$ and $U$, which obey the relations
\begin{eqnarray}
S^2=T^3=U^2=(ST)^3=(SU)^2=(TU)^2=(STU)^4=1\,.
\end{eqnarray}
The $S_4$ group has five irreducible representations: $\mathbf{1}$, $\mathbf{1}^{'}$, $\mathbf{2}$, $\mathbf{3}$ and $\mathbf{3}^{'}$. The one-dimensional unitary representations are given by,
\begin{equation}
\begin{aligned}
&\mathbf{1}~:~ S=1,\quad T=1,\quad U=1\,,\\
&\mathbf{1}^{'}~:~ S=1,\quad T=1,\quad U=-1\,.
\end{aligned}
\end{equation}
In the double representation, we have
\begin{equation}
S=\begin{pmatrix}
1 ~&~0 \\
0 ~&~1
\end{pmatrix},\quad T=\begin{pmatrix}
\omega ~&~0 \\
0 ~&~\omega^2
\end{pmatrix},\quad U=\begin{pmatrix}
0 ~&~ 1 \\
1 ~&~ 0
\end{pmatrix}\,,
\end{equation}
with $\omega=e^{2\pi i/3}$. For the triplet representation $\mathbf{3}$, in a basis where the element $T$ is diagonal, the generators are
\begin{equation}
S=\frac{1}{3}\begin{pmatrix}
-1 ~&~ 2  ~&~ 2  \\
2  ~&~ -1  ~&~ 2 \\
2  ~&~ 2 ~&~ -1
\end{pmatrix},\quad T=\begin{pmatrix}
1  ~&~  0 ~&~ 0 \\
0  ~&~  \omega^2  ~&~ 0 \\
0  ~&~  0   ~&~  \omega
\end{pmatrix},\quad U=-\begin{pmatrix}
1 ~&~  0  ~&~  0  \\
0  ~&~ 0  ~&~ 1 \\
0 ~&~ 1  ~&~  0
\end{pmatrix}\,.
\end{equation}
In another triplet representation $\mathbf{3}'$, the generator $U$ is simply opposite in sign with to that in the $\mathbf{3}$. We assume that both lepton doublet $L$ and the atmospheric flavon $\phi_{\text{atm}}$ are assigned to $S_4$ triplet $\mathbf{3}$, the solar flavon $\phi_{\text{sol}}$ transforms as $\mathbf{3}'$ while the right-handed Majorana neutrino
$N^c_{\mathrm{atm}}$ is $\mathbf{1}$ and $N^c_{\mathrm{sol}}$ is $\mathbf{1^\prime}$ of $S_4$. The residual flavor symmetry of the charged lepton sector is taken to be $G_{l}=Z^{T}_3$. The most general hermitian matrix $m^{\dagger}_{l}m_l$ invariant under $Z^{T}_3$ is diagonal with generic entries, where $m_{l}$ is the charged lepton mass matrix. As a consequence, the unitary transformation $U_{l}$ is a unit matrix up to permutations and rephasing of column vectors. The atmospheric and solar residual symmetries are $G_{\text{atm}}=Z^{U}_2$ and $G_{\text{sol}}=Z^{SU}_2$ with $X_{\text{atm}}=X_{\text{sol}}=1$, and they uniquely fix the vacuum alignments of $\langle\phi_{\text{atm}}\rangle$ and $\langle\phi_{\text{sol}}\rangle$ as follows
\begin{equation}
\langle\phi_{\text{atm}}\rangle=v_{\text{atm}}\begin{pmatrix}
0  \\
1  \\
-1 \\
\end{pmatrix},\quad \langle\phi_{\text{sol}}\rangle=v_{\text{sol}}\begin{pmatrix}
1  \\
n  \\
2-n \\
\end{pmatrix},
\end{equation}
where both vacuum expectation values $v_{\text{atm}}$ and $v_{\text{sol}}$ are real, and $n$ is a real parameter.

It is interesting to compare the two different residual symmetries here, $G_{\text{atm}}=Z^{U}_2$ and $G_{\text{sol}}=Z^{SU}_2$ of the tri-direct CP approach,
to the semi-direct approach of ~\cite{King:2016yvg}
where there was a common residual symmetry in both the atmospheric and solar neutrino sectors, namely $Z^{SU}_2$.
In the semi-direct approach~\cite{King:2016yvg}, the atmospheric alignment $\langle\phi_{\text{atm}}\rangle\propto(0, 1, -1)^{T}$ could not be uniquely fixed by the residual symmetry $Z^{SU}_2$,
and was achieved through the dynamical terms in the potential of a concrete model. As regards the solar alignment, $Z^{SU}_2$ enforces $\langle\phi_{\text{sol}}\rangle\propto(1, n, 2-n)^{T}$ with $n$ being generally complex in~\cite{King:2016yvg} while $n$ is a real parameter because of the CP symmetry in the
tri-direct CP approach.

Applying the seesaw formula and multiplying $L_3$ by a minus sign, we obtain the effective light neutrino mass matrix\footnote{The effective light neutrino Majorana mass matrix given by the Lagrangian $\mathcal{L}_{\text{eff}}=-\frac{1}{2}\nu_{L_i}(m_{\nu})_{ij}\nu_{L_j}+\text{h.c.}$ in two-component notation.}
\begin{equation}
\label{eq:mnu_LS_fir}m_{\nu}=m_a\begin{pmatrix}
0  ~&~  0  ~&~    0   \\
0  ~&~  1  ~&~    1  \\
0  ~&~  1  ~&~    1
\end{pmatrix}+m_s e^{i\eta}\begin{pmatrix}
1   ~&~  n-2  ~&~  n  \\
n-2  ~&~  (n-2)^2 ~&~ n(n-2)  \\
n   ~&~ n(n-2)  ~&~  n^2
\end{pmatrix}\,.
\end{equation}
As shown in~\cite{King:2016yvg,Ballett:2016yod}, the benchmark values of  $n=3$ and $\eta=2\pi/3$ can give a phenomenologically successful description of neutrino masses and lepton mixing parameters, e.g.
\begin{eqnarray}
\nonumber  && m_{a}=26.691\,\text{meV}, \quad m_{s}=2.673\,\text{meV}, \quad \sin^2\theta_{13}=0.0223\,, \\
\nonumber  && \sin^2\theta_{12}=0.318, \quad \sin^2\theta_{23}=0.488, \quad \delta_{CP}=-0.517\pi, \quad  \beta=-0.402\pi\,,\\
\label{eq:predct_LSA}&& m_1=0\,\text{meV},\quad  m_2=8.563\,\text{meV},\quad m_3=49.993\,\text{meV}, \quad m_{ee}=2.673\,\text{meV}\,,
\end{eqnarray}
where $m_{ee}$ is the effective mass in neutrinoless double beta decay. Notice that the predictions in Eq.~\eqref{eq:predct_LSA} generally receive subleading corrections in a concrete model. Since the charged lepton masses are not constrained in this approach, the lepton mixing matrix is determined up to all possible row permutations. In other words, the neutrino mass matrix $m_{\nu}$ in Eq.~\eqref{eq:mnu_LS_fir} can be multiplied by any permutation matrix from the left and right sides simultaneously. If the second and third rows and columns of $m_{\nu}$ are exchanged~\cite{King:2016yvg,Ballett:2016yod}, we obtain a second form of the Littlest seesaw model with
\begin{equation}
m_{\nu}=m_a\begin{pmatrix}
0  ~&~  0  ~&~    0   \\
0  ~&~  1  ~&~    1  \\
0  ~&~  1  ~&~    1
\end{pmatrix}+m_s e^{i\eta}\begin{pmatrix}
1   ~&~  n   ~&~  n-2  \\
n   ~&~  n^2 ~&~ n(n-2)  \\
n-2  ~&~ n(n-2)  ~&~ (n-2)^2
\end{pmatrix}\,.
\end{equation}
The phase value $\eta=-2\pi/3$ is preferred in the case of $n=3$. Excellent agreement with experimental data can be achieved, and a numerical benchmark is
\begin{eqnarray}
\nonumber && m_{a}=26.688\,\text{meV}, \quad m_{s}=2.674\,\text{meV},  \quad \sin^2\theta_{13}=0.0223\,, \\
\nonumber &&\sin^2\theta_{12}=0.318, \quad \sin^2\theta_{23}=0.512, \quad \delta_{CP}=-0.483\pi, \quad  \beta=0.402\pi\,, \\
&& m_1=0\,\text{meV},\quad  m_2=8.565\,\text{meV},\quad m_3=49.991\,\text{meV}, \quad m_{ee}=2.674\,\text{meV}\,.
\end{eqnarray}
For both versions of Littlest seesaw, the neutrino mass matrix leads to the trimaximal TM1 mixing~\cite{King:2015dvf,King:2016yvg,Ballett:2016yod}, in which the first column of the mixing matrix is fixed to be that of the tri-bimaximal mixing matrix. See~\cite{King:2015dvf,King:2016yvg} for exact expressions of light neutrino masses and lepton mixing parameters.

\section{\label{sec:new_LSS}Littlest Seesaw variants in the tri-direct CP approach}

Following the framework of tri-direct CP approach presented in section~\ref{sec:framework}, we find that many new mixing patterns compatible with experimental data can obtained from the $S_4$ flavor symmetry group and CP~\cite{Ding:2018tuj}. In order to illustrate how the neutrino mass and mixing parameters are predicted in the tri-direct CP approach, as an example, we shall show a new Littlest seesaw model which can be achieved from the $S_4$ flavor symmetry in combination with the generalized CP symmetry $H_{CP}$, where $H_{CP}$ is the collection of the generalized CP transformations $X_{\bf r}$. In our working basis, the generalized CP transformation compatible with $S_{4}$ turns out to be of the same form as flavor symmetry transformation~\cite{Ding:2013hpa,Li:2013jya}, i.e.
\begin{equation}
X_{\bf{r}}=\rho_{\mathbf{r}}(h), \qquad h\in S_{4}\,,
\end{equation}
where $h$ can be any element of $S_4$. The three left-handed leptons $L$ are assigned to $S_4$ triplet $\mathbf{3}$, the atmospheric flavon $\phi_{\text{atm}}$ and solar flavon $\phi_{\text{sol}}$ transform as $\mathbf{3}$ and $\mathbf{3}'$ respectively, and the two right-handed neutrinos $N^c_{\mathrm{atm}}$ and $N^c_{\mathrm{sol}}$ are $S_4$ singlets $\mathbf{1}$ and $\mathbf{1}'$ respectively. We assume that the $S_4$ and CP symmetries are broken down to $Z^T_3$, $Z^{TST^2}_2\times H^{\text{atm}}_{CP}$ and $Z_2^U\times H^{\text{sol}}_{CP}$ in the charged lepton, atmospheric neutrino and solar neutrino sectors, respectively. Since the representation matrix $T$ is diagonal in the working basis, the residual symmetry $G_{l}=Z^T_3$ would determine that the hermitian combination $m^{\dagger}_lm_{l}$ is diagonalized by a unit matrix up to permutations and phases of its column vectors. Notice that no new constraint is obtained even if the possible residual CP symmetry in the charged lepton sector is further taken into account~\cite{Yao:2016zev}.

The residual symmetry $Z^{TST^2}_2\times H^{\text{atm}}_{CP}$ in the atmospheric neutrino sector should be well defined such that the constrained consistency condition in Eq.~\eqref{eq:nu_atm_consistency} has to be satisfied, i.e.
\begin{equation}
X^{\text{atm}}_{\mathbf{r}}\rho^{*}_{\mathbf{r}}(TST^2)(X^{\text{atm}}_{\mathbf{r}})^{-1}=\rho_{\mathbf{r}}(TST^2)\,,
\end{equation}
It is easy to check that $X^{\text{atm}}_{\mathbf{r}}$ can only take the following eight possible values,
\begin{small}
\begin{equation}\label{eq:CP_atm}
H^{\text{atm}}_{CP}=\{\rho_{\mathbf{r}}(SU), \rho_{\mathbf{r}}(T^2), \rho_{\mathbf{r}}(ST^2S), \rho_{\mathbf{r}}(T^2STU),
\rho_{\mathbf{r}}(U),\rho_{\mathbf{r}}(ST^2), \rho_{\mathbf{r}}(T^2S),\rho_{\mathbf{r}}(TST^2U)\}\,.
\end{equation}
\end{small}
For the first four CP transformations of $H^{\text{atm}}_{CP}$, the vacuum alignment $\langle\phi_{\text{atm}}\rangle$ which preserves  $Z^{TST^2}_2\times H^{\text{atm}}_{CP}$ is of the following form
\begin{equation}\label{eq:VEV_atm}
\langle\phi_{\text{atm}}\rangle=v_{\text{atm}}\left(1, \omega^2, \omega\right)^T\,,
\end{equation}
where the parameter $v_{\text{atm}}$ is real. If $X^{\text{atm}}_{\mathbf{r}}$ is one of the last four CP transformations in $H^{\text{atm}}_{CP}$ (e.g. $X_{\text{atm}}=U$), the VEV $\langle\phi_{\text{atm}}\rangle$ is enforced to be the form
\begin{equation}\label{eq:VEV_atm2}
\langle\phi_{\text{atm}}\rangle=iv_{\text{atm}}\left(1, \omega^2, \omega\right)^T\,,
\end{equation}
with $v_{\text{atm}}$ being real. The above two vacuum configurations in Eq.~\eqref{eq:VEV_atm} and Eq.~\eqref{eq:VEV_atm2} differ from each other in the overall factor $i$ which can be compensated by shifting the sign of the coupling $x_{\text{atm}}$. Without loss of generality, in the following we shall take the atmospheric vacuum in Eq.~\eqref{eq:VEV_atm} and the corresponding residual CP is $X_{\text{atm}}=SU$. As regards the solar neutrino sector. the residual CP transformation $X^{\text{sol}}_{\bf{r}}$ of $H^{\text{sol}}_{CP}$ is determined by the consistency condition
\begin{equation}
X^{\text{sol}}_{\mathbf{r}}\rho^{*}_{\mathbf{r}}(U)(X^{\text{sol}}_{\mathbf{r}})^{-1}
=\rho_{\mathbf{r}}(U)\,.
\end{equation}
We find the allowed values of $H^{\text{sol}}_{CP}$ are
\begin{equation}\label{eq:CP_sol}
H^{\text{sol}}_{CP}=\{\rho_{\bf r}(1), \rho_{\bf r}(U), \rho_{\bf r}(S), \rho_{\bf r}(SU)\}\,.
\end{equation}
For the case of $X^{\text{sol}}_{\bf{r}}=\rho_{\mathbf{r}}(1), \rho_{\mathbf{r}}(U)$, the vacuum configuration $\langle\phi_{\text{sol}}\rangle$ invariant under $Z_2^U\times H^{\text{sol}}_{CP}$ is of the form
\begin{equation}\label{eq:VEV_sol}
\langle\phi_{\text{sol}}\rangle=v_{\text{sol}}\left(1, x, x\right)^T\,,
\end{equation}
where the VEV $v_{\text{sol}}$ is real and $x$ is in general a dimensionless real number. For the remaining choices $X^{\text{sol}}_{\bf{r}}=\rho_{\bf r}(S), \rho_{\bf r}(SU)$, the subgroup $Z_2^U\times H^{\text{sol}}_{CP}$ is preserved only if $\langle\phi_{\text{sol}}\rangle$ is aligned along the following direction,
\begin{equation}\label{eq:VEV_sol_2}
\langle\phi_{\text{sol}}\rangle=v_{\text{sol}}\left(1+2ix, 1-ix,1-ix\right)^T\,,
\end{equation}
with both parameters $v_{\text{sol}}$ and $x$ being real. However, detailed numerical analysis reveals that the experimentally favored neutrino masses and mixing angles can not be obtained for the solar flavon VEV shown in Eq.~\eqref{eq:VEV_sol_2}. Hence we shall focus on the residual CP transformation $X_{\text{sol}}=U$ and the resulting vacuum alignment $\langle\phi_{\text{sol}}\rangle\propto(1,x, x)$ in this work.

\begin{figure}[t!]
\centering
\begin{tabular}{cc}
\includegraphics[width=0.7\linewidth]{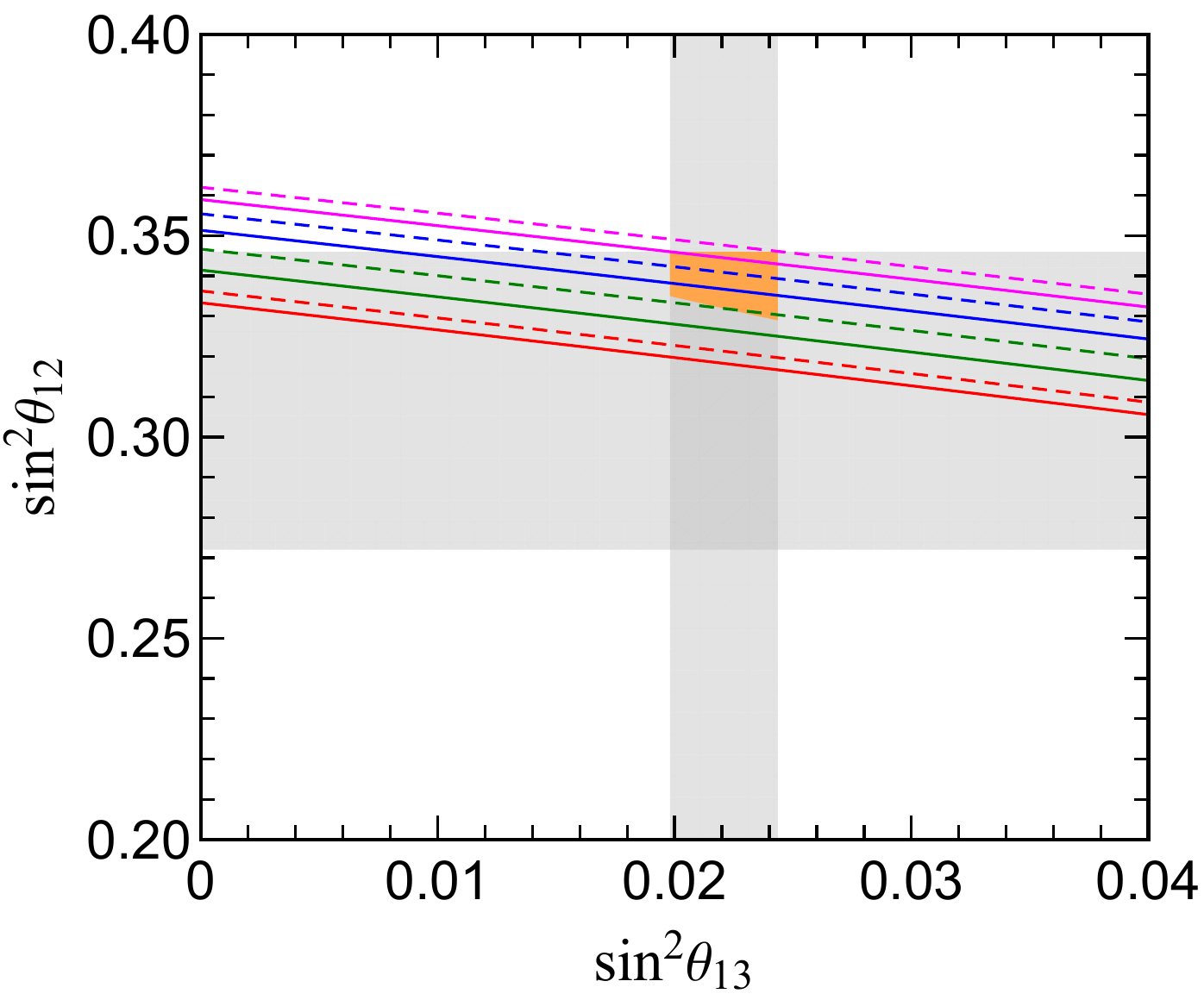}&
\multirow{1}[30]{0.5\linewidth}[8.0cm]{\includegraphics[width=0.5\linewidth]{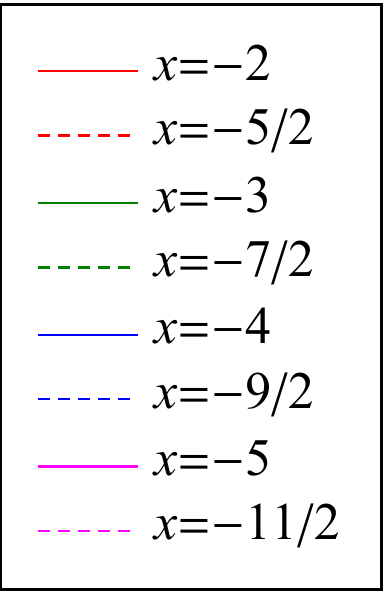}}
\end{tabular}
\caption{\label{fig:contour_n} Correlation between $\sin^2\theta_{12}$ and $\sin^2\theta_{13}$ given by Eq.~\eqref{eq:correlation_mix_angles} for various values of $x$. The gray bands represent the experimentally preferred $3\sigma$ ranges of $\sin^2\theta_{13}$ and $\sin^2\theta_{12}$ adapted from~\cite{Esteban:2016qun}. The orange area denotes the most generally allowed regions of $\sin^2\theta_{13}$ and $\sin^2\theta_{12}$ in the new Littlest seesaw model, where the four input parameter $m_a$, $m_s$, $\eta$ and $x$ are randomly chosen and the resulting mixing angles and mass squared differences are required to lie in the experimentally preferred $3\sigma$ regions~\cite{Esteban:2016qun}.}
\end{figure}

It is useful to remind the $S_4$ singlet contraction rules for $\mathbf{3}\otimes\mathbf{3}\rightarrow\mathbf{1}$ and $\mathbf{3}\otimes\mathbf{3}'\rightarrow\mathbf{1}'$~\cite{Ding:2013hpa},
\begin{equation}
\begin{aligned}
\left(\alpha\beta\right)_{\mathbf{1}}=\alpha_1\beta_1+\alpha_2\beta_3+\alpha_3\beta_2\,,\\
\left(\alpha\gamma\right)_{\mathbf{1}'}=\alpha_1\gamma_1+\alpha_2\gamma_3+\alpha_3\gamma_2\,,
 \end{aligned}
\end{equation}
where $\alpha=(\alpha_1, \alpha_2, \alpha_3)^{T}$ and $\beta=(\beta_1, \beta_2, \beta_3)^{T}$ are $S_4$ triplet $\mathbf{3}$, and $\gamma=(\gamma_1, \gamma_2, \gamma_3)^{T}$ transforms as $\mathbf{3}'$. When the flavor fields $\phi_{\text{atm}}$ and $\phi_{\text{sol}}$ acquire a non-vanishing VEV as shown in Eq.~\eqref{eq:VEV_atm} and Eq.~\eqref{eq:VEV_sol} respectively, from the Lagrangian of Eq.~\eqref{eq:Lagrangian}, we can read out the Dirac neutrino mass matrix $m_D$ as well as a diagonal right-handed neutrino mass matrix $m_{N}$
\begin{equation}
m_{D}=\begin{pmatrix}
y_{\text{atm}}v_{\text{atm}}      ~&~    y_{\text{sol}}v_{\text{sol}}  \\
\omega y_{\text{atm}}v_{\text{atm}}   ~&~  x y_{\text{sol}}v_{\text{sol}} \\
\omega^2y_{\text{atm}}v_{\text{atm}}  ~&~  x y_{\text{sol}}v_{\text{sol}}
\end{pmatrix}, \qquad m_{N}=\begin{pmatrix}
M_{\textrm{atm}}  &  0  \\
0  &   M_{\textrm{sol}}
\end{pmatrix}\,,
\end{equation}
with $M_{\textrm{atm}}=x_{\textrm{atm}}\langle\xi_{\text{atm}}\rangle$ and $M_{\textrm{sol}}=x_{\textrm{sol}}\langle\xi_{\text{sol}}\rangle$.
The light neutrino mass matrix is given by the seesaw formula,
\begin{equation}
\label{eq:mnu}  m_{\nu}=-m_{D}m^{-1}_{N}m^{T}_{D} =m_{a}\begin{pmatrix}
 1 &~ \omega  &~ \omega ^2 \\
 \omega  &~ \omega ^2 &~ 1 \\
 \omega ^2 &~ 1 &~ \omega  \\
\end{pmatrix}+e^{i\eta}m_{s}
\begin{pmatrix}
 1 &~  x &~  x \\
 x &~ x^2 &~ x^2 \\
 x &~ x^2 &~ x^2 \\
\end{pmatrix}\,,
\end{equation}
where a physically irrelevant overall phase factor is dropped, $m_{a}=|y^2_{\text{atm}}v^2_{\text{atm}}/M_{\textrm{atm}}|$, $m_{s}=|y^2_{\text{sol}}v^2_{\text{sol}}/M_{\mathrm{sol}}|$ and the relative phase $\eta=2\mathrm{arg}(y_{\text{sol}}v_{\text{sol}})-2\mathrm{arg}(y_{\text{atm}}v_{\text{atm}})+\mathrm{arg}(M_{\text{atm}})-\mathrm{arg}(M_{\text{sol}})$. We see that $m_{\nu}$ depends on four parameters $m_a$, $m_s$, $\eta$ and $x$ to describe the neutrino masses and mixing parameters, consequently this model is quite predictive. Moreover, we shall fix $x$ and $\eta$ to certain values in a concrete models. It is easy to check that the above neutrino mass matrix $m_{\nu}$ fulfills
\begin{equation}
m_{\nu}\begin{pmatrix}
-i \sqrt{3} x \\
x-\omega^2 \\
-x+\omega
\end{pmatrix}=\left(\begin{array}{c}
0\\
0\\
0
\end{array}\right)\,.
\end{equation}
This implies that the column vector $(-i\sqrt{3} x, x-\omega^2, -x+\omega)^{T}$ is an eigenvector of $m_\nu$ with zero eigenvalue. As a consequence, the neutrino mass matrix $m_{\nu}$ can be block diagonalized by the following unitary transformation
\begin{equation}
U_{\nu1}=\begin{pmatrix}
 -\frac{i \sqrt{3} x}{\sqrt{5x^2+2x+2}} &~ \sqrt{\frac{2 \left(x^2+x+1\right)}{5x^2+2x+2}} &~ 0 \\
 \frac{x-\omega^2}{\sqrt{5x^2+2x+2}} &~ -\frac{i \sqrt{3}x\left(x-\omega^2\right)}{\sqrt{2\left(x^2+x+1\right) \left(5x^2+2x+2\right)}} &~ \frac{x-\omega^2}{\sqrt{2(x^2+x+1)}} \\
 \frac{\omega-x}{\sqrt{5x^2+2x+2}} &~ \frac{i \sqrt{3}x (x-\omega)}{\sqrt{2 \left(x^2+x+1\right) \left(5x^2+2x+2\right)}} &~ \frac{x-\omega}{\sqrt{2(x^2+x+1)}}
\end{pmatrix}\,.
\end{equation}
Then $m_{\nu}$ becomes
\begin{equation}\label{eq:nu_mass_matrix2}
m^\prime_{\nu}=U^T_{\nu1}m_{\nu}U_{\nu1}=\begin{pmatrix}
0  &~   0  &~  0 \\
0  &~  y   &~ z  \\
0  &~  z  &~ w
\end{pmatrix}\,,
\end{equation}
with
\begin{eqnarray}
\nonumber && y=\frac{5x^2+2x+2}{2\left(x^2+x+1\right)}(m_{a}+e^{i \eta } m_{s})\equiv|y|e^{i\phi_y}\,, \\
\nonumber && z=-\frac{\sqrt{5x^2+2x+2}}{2\left(x^2+x+1\right)}\left[ (x+2)m_{a}-x(2x+1)e^{i \eta }m_{s}\right]\equiv|z|e^{i\phi_z}\,, \\
\label{eq:yzw_par}&& w=\frac{1}{2(x^2+x+1)}\left[(x+2)^2m_{a}+x^2\left(2x+1\right)^2e^{i \eta } m_{s}\right]\equiv|w|e^{i\phi_w}\,.
\end{eqnarray}
Since $m'_{\nu}$ is essentially a two by two complex symmetric matrix, it can be exactly diagonalized, as shown in detail in~\cite{Ding:2013bpa},
\begin{equation}
U^{T}_{\nu2}m^{\prime}_{\nu}U_{\nu2}=\text{diag}(0, m_2, m_3) \,.
\end{equation}
The procedure for diagonalization of the neutrino mass matrix $m^{\prime}_{\nu}$ is given in Appendix~\ref{sec:appendix_A}, and the explicit forms of $U_{\nu2}$, $m_{2}$ and $m_{3}$ can be found there. Because the charged lepton mass matrix $m^{\dagger}_lm_{l}$ is a diagonal, the lepton mixing completely arises from the neutrino sector. Thus the lepton mixing matrix is derived as
\begin{eqnarray}
\nonumber&&\hskip-0.3in U_{PMNS}=U_{\nu1}U_{\nu2}\\
&&\hskip-0.1in \quad =\frac{1}{\sqrt{2}}\begin{pmatrix}
 \frac{ \sqrt{6} x}{\sqrt{5x^2+2x+2}} &~  2i\sqrt{\frac{x^2+x+1}{5x^2+2x+2}} \cos \theta  &~ 2i\sqrt{\frac{x^2+x+1}{5x^2+2x+2}} e^{i \psi }\sin \theta  \\
 \sqrt{\frac{2(x^2+x+1)}{5x^2+2x+2}} &~ -e^{-i \psi } \sin \theta -\frac{i \sqrt{3} x \cos\theta }{\sqrt{5x^2+2x+2}} &~ \cos \theta -\frac{i \sqrt{3}x e^{i \psi }  \sin \theta }{\sqrt{5x^2+2x+2}} \\
 \sqrt{\frac{2(x^2+x+1)}{5x^2+2x+2}} &~ e^{-i \psi } \sin \theta -\frac{i \sqrt{3}x\cos \theta }{\sqrt{5x^2+2x+2}} &~ -\cos \theta -\frac{i \sqrt{3}x e^{i \psi}\sin\theta}{\sqrt{5x^2+2x+2}} \\
\end{pmatrix}P_{\nu}\,,
\end{eqnarray}
up to possible row permutations, where $P_{\nu}=\text{diag}(1, e^{i(\psi+\rho)/2}, e^{i(-\psi+\sigma)/2})$ is a diagonal phase matrix, and an overall phase of each row has been absorbed by the charged lepton fields. The expressions for the parameters $\theta$, $\psi$, $\rho$ and $\sigma$ are given in Appendix~\ref{sec:appendix_A}. Comparing with the standard parametrization of the lepton mixing matrix~\cite{Patrignani:2016xqp},
\begin{equation}\label{eq:PMNS_def}
U_{PMNS}=\left(\begin{array}{ccc}
c_{12}c_{13}  &   s_{12}c_{13}   &   s_{13}e^{-i\delta_{CP}}  \\
-s_{12}c_{23}-c_{12}s_{13}s_{23}e^{i\delta_{CP}}   &  c_{12}c_{23}-s_{12}s_{13}s_{23}e^{i\delta_{CP}}  &  c_{13}s_{23}  \\
s_{12}s_{23}-c_{12}s_{13}c_{23}e^{i\delta_{CP}}   & -c_{12}s_{23}-s_{12}s_{13}c_{23}e^{i\delta_{CP}}  &  c_{13}c_{23}
\end{array}\right)\text{diag}(e^{i\frac{\alpha}{2}},e^{i\frac{\beta}{2}},1)\,,
\end{equation}
where $c_{ij}\equiv \cos\theta_{ij}$, $s_{ij}\equiv \sin\theta_{ij}$, and the Majorana phase $\alpha$ is unphysical since $m_1=0$, we can extract the results for the lepton mixing angles and find
\begin{eqnarray}
\nonumber \sin^2\theta_{13}&=&\frac{2\left(x^2+x+1\right)\sin^2\theta}{5x^2+2x+2}\,,\\
\nonumber \sin^2\theta_{12}&=&1-\frac{3x^2 }{3x^2+2\left(x^2+x+1\right) \cos^2\theta }\,,\\
\label{eq:angles}\sin^2\theta_{23}&=&\frac{1}{2}+\frac{x\sqrt{3\left(5x^2+2x+2\right)}  \sin2\theta\sin\psi }{2\left[3x^2+2\left(x^2+x+1\right) \cos^ 2 \theta\right]}\,,
\end{eqnarray}
and for the CP invariants we obtain
\begin{equation}
J_{CP}=\frac{\sqrt{3} x\left(x^2+x+1\right)\sin2\theta\cos\psi}{2\left(5 x^2+2x+2\right)^{3/2}}\,, \quad
I_{1}=\frac{\left(x^2+x+1\right)^2\sin^22\theta\sin (\rho -\sigma )}{\left(5x^2+2x+2\right)^2} \,.
\end{equation}
The Jarlskog invariant $J_{CP}$~\cite{Jarlskog:1985ht} and the Majorana invariant $I_1$~\cite{Branco:1986gr,Nieves:1987pp,Jenkins:2007ip,Branco:2011zb} related to the Majorana phase $\beta$ are defined in the usual way
\begin{eqnarray}
\nonumber && J_{CP}=\Im{(U_{PMNS,11}U_{PMNS,33}U^{*}_{PMNS,13}U^{*}_{PMNS,31})}=\frac{1}{8}\sin2\theta_{12}\sin2\theta_{13}\sin2\theta_{23}\cos\theta_{13}\sin\delta_{CP}\,, \\
\label{eq:CP_invariants}&& I_{1}=
\Im{(U^{2}_{PMNS, 12}U^{*2}_{PMNS, 13})}=\frac{1}{4}\sin^2\theta_{12}\sin^22\theta_{13}\sin(\beta+2\delta_{CP})\,.
\end{eqnarray}
From the expressions of mixing angles in Eq.~\eqref{eq:angles}, we easily see that the solar mixing angle $\theta_{12}$ and the reactor mixing angle $\theta_{13}$ fulfill the following sum rules
\begin{equation}\label{eq:correlation_mix_angles}
\cos^2\theta_{12}\cos^2\theta_{13}=\frac{3x^2}{5 x^2+2x+2}\,.
\end{equation}
This implies that $\theta_{13}$ and $\theta_{12}$ are strongly correlated with each other, as shown in figure~\ref{fig:contour_n}.
Inserting the $3\sigma$ allowed regions $0.272\leq\sin^2\theta_{12}\leq0.346$ and $0.01981\leq\sin^2\theta_{13}\leq0.02436$~\cite{Esteban:2016qun}, we find that the parameter $x$ should be in the interval $-5.481\leq x\leq-1.223$. Notice that the value of $x$ is also subject to the constraints from the measured values of $\theta_{23}$ and neutrino mass squared differences. If all the four input parameters $m_a$, $m_s$, $\eta$ and $x$ are treated as free parameters and both lepton mixing angles and the mass splittings $\Delta m^2_{21}$ and $\Delta m^2_{31}$ are required to be in the experimentally favored $3\sigma$ ranges, we find the solar mixing angle is allowed in a narrow region $0.329\leq\sin^2\theta_{12}\leq0.346$ which is represented by the orange in figure~\ref{fig:contour_n}. The forthcoming JUNO experiment will be capable of reducing the error of $\sin^2\theta_{12}$ to about $0.1^{\circ}$ or around $0.3\%$~\cite{An:2015jdp}. Therefore we can expect JUNO to identify with considerable confidence if the present model is compatible with experimental data.

\begin{figure}[t!]
\centering
\begin{tabular}{c}
\includegraphics[width=1.0\linewidth]{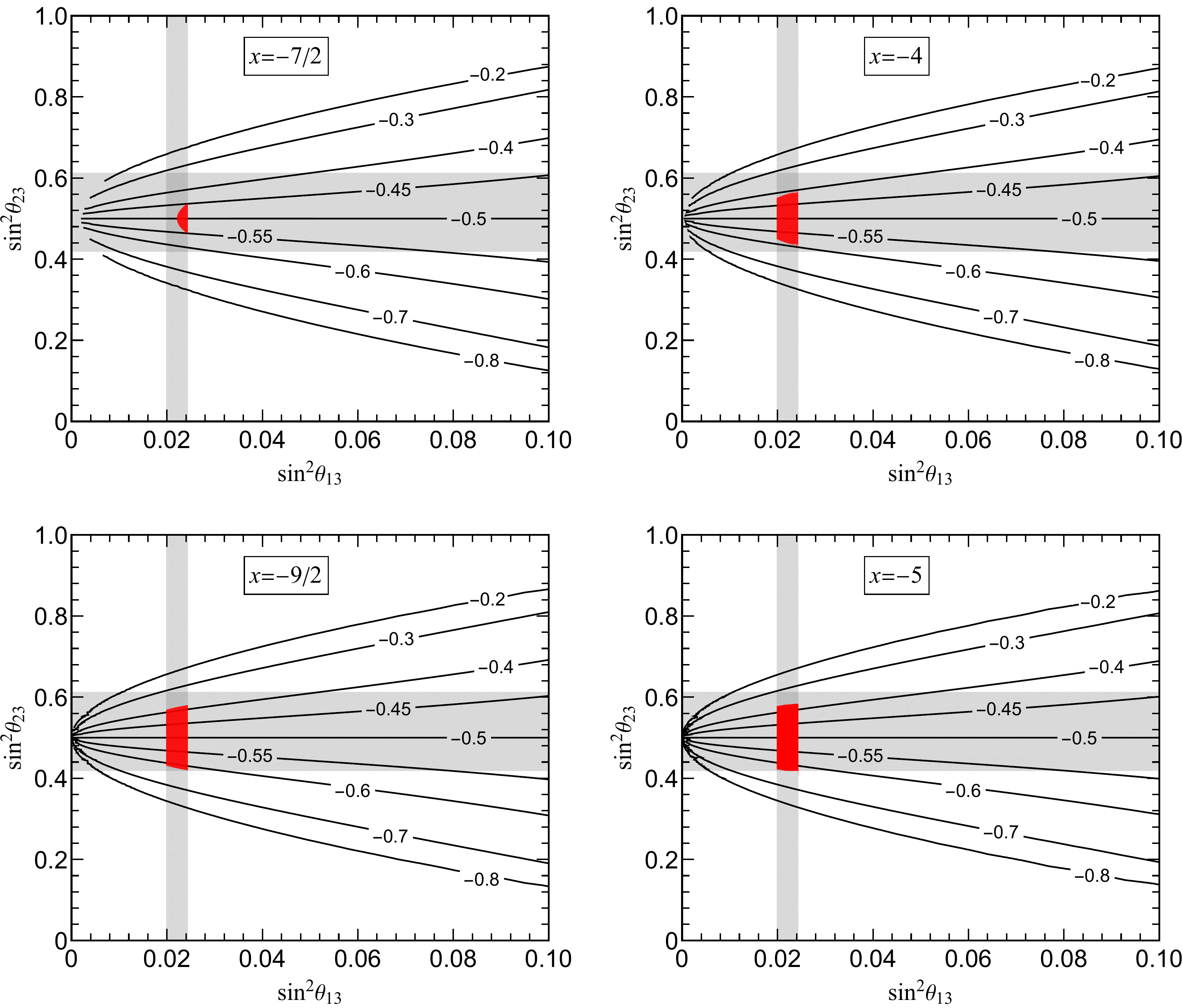}
\end{tabular}
\caption{\label{fig:contour_CP} Contour plot of $\delta_{CP}/\pi$ in the $\sin^2\theta_{13}-\sin^2\theta_{23}$ plane for the benchmark values of $x=-7/2, -4, -9/2, -5$. The contour lines are obtained by using the sum rule in Eq.~\eqref{eq:delta_CP_sumrule}. The gray bands represent the experimentally preferred $3\sigma$ ranges of the mixing angles adapted from~\cite{Esteban:2016qun}. The red areas in the plane are the most generally allowed regions of $\sin^2\theta_{13}$ and $\sin^2\theta_{23}$ for given value of $x$, where the four input parameter $m_a$, $m_s$ and $\eta$ are randomly chosen and the resulting mixing angles and mass squared differences are required to lie in the experimentally preferred $3\sigma$ regions~\cite{Esteban:2016qun}.}
\end{figure}

It is notable that all the three mixing angles and Jarlskog invariant $J_{CP}$ depend on only two parameters $\theta$ and $\psi$. As a consequence, we can express the Dirac CP phase $\delta_{CP}$ in terms of the mixing angles,
\begin{eqnarray}
\nonumber \cos\delta_{CP}&=&\frac{ \cot 2 \theta_{23} \left[3x^2-\left(4x^2+ x+1\right)\cos^2\theta_{13}\right]}{\sqrt{3} \left|x\right| \sin \theta_{13} \sqrt{\left(5x^2+2x+2\right)\cos^2\theta_{13}-3x^2}}\,,\\
\label{eq:delta_CP_sumrule}\sin\delta_{CP}&=& \text{sign}(x\cos\psi)\csc 2 \theta_{23} \sqrt{1+\frac{\left(x^2+x+1\right)^2 \cot ^2\theta_{13} \cos ^22 \theta_{23}}{3x^2 \left[3x^2 \tan ^2\theta_{13}-2 \left(x^2+x+1\right)\right]}}~\,,
\end{eqnarray}
For maximal atmospheric mixing angle $\theta_{23}=\pi/4$, we have $\cos2\theta_{23}=0$ and $\csc2\theta_{23}=1$. Then this sum rule gives $\cos\delta_{CP}=0$ and $\sin\delta_{CP}=\pm1$ which implies maximal Dirac phase $\delta_{CP}=\pm\pi/2$. We show the contour plot of $\delta_{CP}/\pi$ in the plane $\sin^2\theta_{23}$ versus $\sin^2\theta_{13}$ in figure~\ref{fig:contour_CP} for $x=-7/2, -4, -9/2, -5$. It is remarkable that the Dirac CP violation phase $\delta_{CP}$ is predicted to lie in a narrow range around $-0.5\pi$.

Furthermore we perform a comprehensive numerical analysis. The input parameters $x$, $r=m_s/m_a$ and $\eta$ are treated as random real numbers in the ranges $x\in[-6, -1]$, $r\in[0, 10]$ and $\eta\in[0, 2\pi]$, then we calculate the values of mixing angles $\sin^2\theta_{ij}$, the CP violation phases $\delta_{CP}$ and $\beta$ and the mass ratio $m^2_2/m^2_3$ for each value of the input parameters $x$, $r$ and $\eta$. We require $\sin^2\theta_{ij}$ and $m^2_2/m^2_3$ to lie in their $3\sigma$ regions obtained in the global analysis of neutrino oscillation data~\cite{Esteban:2016qun}. In order to accommodate the present experimental data, we find the allowed region of the parameter $x$ is $-5.475\leq x\leq-3.370$. Regarding the predictions for the mixing angles, we find that all values of $\sin^2\theta_{13}$ in its $3\sigma$ range are allowed, $\sin^2\theta_{23}$ is constrained to lie in the interval $0.418\leq\sin^2\theta_{23}\leq0.584$, and the solar angle is found to lie in a narrow interval around its $3\sigma$ upper bound $0.329\leq\sin^2\theta_{12}\leq0.346$. What concerns the CP phases, the values of $\delta_{CP}$ lie around $-\pi/2$, in the range $-0.629\pi\leq\delta_{CP}\leq-0.371\pi$, and the allowed range of the Majorana phase is $-0.571\pi\leq\beta\leq0.571\pi$. These predictions may be tested at future long baseline experiments, as discussed in~\cite{Ballett:2016yod}.

In order to quantitatively measure how well the present model can describe the experimental data, we define a $\chi^2$ function to estimate the goodness-of-fit of a chosen values of the input parameters $m_a$, $r$, $\eta$ and $x$,
\begin{equation}
\chi^2 = \sum_{i=1}^5 \left( \frac{P_i(m_a, r,\eta, x)-O_i}{\sigma_i}\right)^2\,,
\label{eq:chisq}
\end{equation}
where $O_i\in \{\sin^2\theta_{12}, \sin^2\theta_{23}, \sin^2\theta_{13}, \Delta m_{21}^2,  \Delta m_{31}^2\}$ are the global best fit values of the observable quantities, and $\sigma_i$ denote the $1\sigma$ deviations of
the corresponding quantities. The values of $O_i$ and $\sigma_i$ are taken from the global data analysis~\cite{Esteban:2016qun}. $P_i$ is the theoretical predictions for the mixing angles $\sin^2\theta_{ij}$ and the mass splittings $\Delta m^2_{21}$ and $\Delta m^2_{31}$ as complex nonlinear
functions of the free parameters of the model. Notice that the value of Dirac phase $\delta_{CP}$ is less constrained at present, consequently its contribution is not included in the $\chi^2$ function. Since the number of free parameters is less than the number of observables, it is not completely evident that the model can successfully fit the data. For each value of the input parameters, we can obtain the predicted values $P_i$ and the corresponding $\chi^2$, and the optimum input parameters yield the lowest $\chi^2$. We have carried out the $\chi^2$ minimization, we find the minimum of $\chi^2$ is $\chi^2_{\text{min}}=3.957$, and values of the input parameters at the $\chi^2_{\text{min}}$ read
\begin{equation}
m_{a}=3.709\,\text{meV}, \quad r=0.537, \quad \eta=1.055\pi,\quad  x=-3.556\,.
\end{equation}
The predictions for various observable quantities obtained at the best fit point are
\begin{eqnarray}
\nonumber &&\sin^2\theta_{13}=0.0222, \quad \sin^2\theta_{12}=0.332, \quad \sin^2\theta_{23}=0.515, \\
\nonumber&& \delta_{CP}=-0.478\pi,\quad \beta=0.0574\pi,\quad  m_1=0\,\text{meV} \,, \\
&&m_2=8.604\,\text{meV}, \quad m_3=49.938\,\text{meV}, \quad m_{ee}=1.778\,\text{meV}\,.
\end{eqnarray}

\begin{figure}[t!]
\centering
\begin{tabular}{c}
\includegraphics[width=0.7\linewidth]{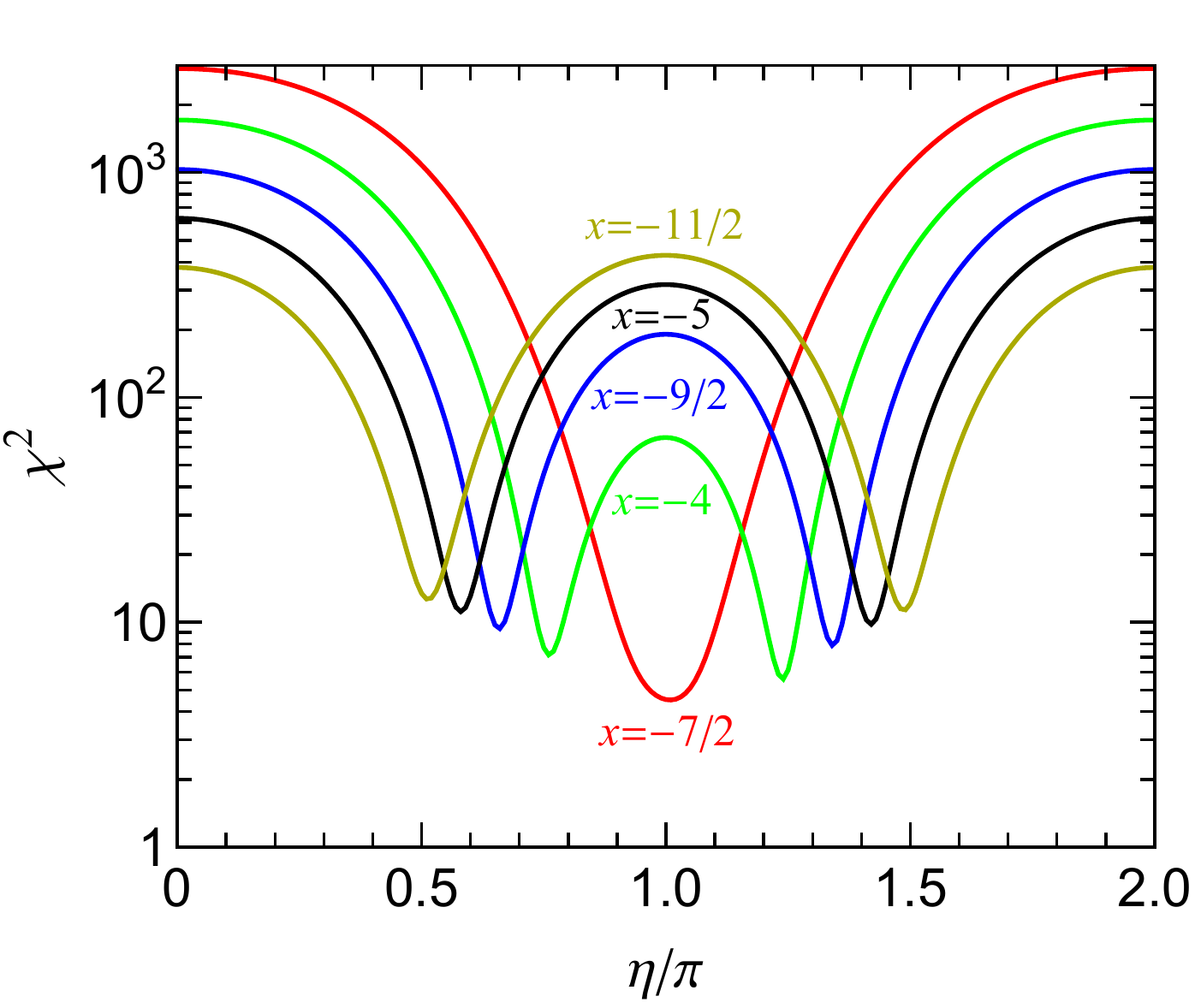}
\end{tabular}
\caption{\label{fig:correlation} Variation of $\chi^2$ with respect to the phase $\eta$ for the typical values of $x=-7/2, -4, -9/2, -5, -11/2$.}
\end{figure}

\begin{table}[t!]
\renewcommand{\tabcolsep}{0.7mm}
\renewcommand{\arraystretch}{1.3}
\small
\centering
\begin{tabular}{|c|c| c| c| c | c| c| c| c| c| c|c |c|c|c|}  \hline \hline
 $x$   & $\eta$  & $m_{a}(\text{meV})$ & $r$ 	 & $\chi^2_{\text{min}}$ &  $\sin^2\theta_{13}$  &$\sin^2\theta_{12}$  & $\sin^2\theta_{23}$  & $\delta_{CP}/\pi$ &  $\beta/\pi$ & $m_2(\text{meV})$ & $m_3(\text{meV})$ & $m_{ee}(\text{meV})$ \\   \hline
 \multirow{6}{*}{$-4$} & $\frac{3 \pi }{4}$ & $3 .723$ & $0 .419$ & $7 .703$ & $0 .0226$ & $0 .336$ & $0 .440$ & $-0.588$ & $-0.264$ & $8 .603$ & $49 .939$ & $2 .843$ \\ \cline{2-15}
 & $\frac{5\pi }{4}$ & $3 .723$ & $0 .419$ & $6 .130$ & $0 .0226$ & $0 .336$ & $0 .560$ & $-0.412$ & $0 .264$ & $8 .603$ & $49 .939$ & $2 .843$  \\ \cline{2-15}
 & $\frac{4\pi }{5}$ & $3 .690$ & $0 .425$ & $12 .206$ & $0 .0204$ & $0 .338$ & $0 .450$ & $-0.577$ & $-0.211$ & $8 .577$ & $49 .974$ & $2 .591$ \\ \cline{2-15}
 & $\frac{6\pi }{5}$ & $3 .691$ & $0 .425$ & $10 .716$ & $0 .0204$ & $0 .338$ & $0 .550$ & $-0.423$ & $0 .211$ & $8 .578$ & $49 .973$ & $2 .591$ \\ \cline{2-15}
& $\frac{7 \pi }{9}$ & $3 .706$ & $0 .422$ & $8 .135$ & $0 .0213$ & $0 .337$ & $0 .446$ & $-0.583$ & $-0.234$ & $8 .592$ & $49 .954$ & $2 .701$ \\ \cline{2-15}
 & $\frac{11 \pi }{9}$ & $3 .706$ & $0 .422$ & $6 .588$ & $0 .0213$ & $0 .337$ & $0 .555$ & $-0.417$ & $0 .234$ & $8 .593$ & $49 .953$ & $2 .701$  \\ \hline
\multirow{6}{*}{$-5$} & $\frac{3 \pi }{5}$ & $3 .737$ & $0 .264$ & $13 .077$ & $0 .0211$ & $0 .345$ & $0 .423$ & $-0.622$ & $-0.423$ & $8 .625$ & $49 .914$ & $3 .558$ \\ \cline{2-15}
 & $\frac{7 \pi }{5}$ & $3 .736$ & $0 .264$ & $11 .693$ & $0 .0211$ & $0 .345$ & $0 .577$ & $-0.378$ & $0 .422$ & $8 .623$ & $49 .916$  & $3 .557$ \\ \cline{2-15}
 & $\frac{4 \pi }{7}$ & $3 .738$ & $0 .262$ & $11 .547$ & $0 .0226$ & $0 .344$ & $0 .422$ & $-0.619$ & $-0.452$ & $8 .586$ & $49 .962$  & $3 .647$ \\ \cline{2-15}
 & $\frac{10 \pi }{7}$ & $3 .737$ & $0 .263$ & $10 .201$ & $0 .0226$ & $0 .344$ & $0 .578$ & $-0.381$ & $0 .452$ & $8 .585$ & $49 .964$  & $3 .646$ \\ \cline{2-15}
 & $\frac{5 \pi }{9}$ & $3 .736$ & $0 .262$ & $14 .269$ & $0 .0234$ & $0 .344$ & $0 .421$ & $-0.618$ & $-0.469$ & $8 .559$ & $49 .999$ & $3 .694$ \\ \cline{2-14}
 & $\frac{13 \pi }{9}$ & $3 .735$ & $0 .262$ & $12 .937$ & $0 .0234$ & $0 .344$ & $0 .579$ & $-0.382$ & $0 .469$ & $8 .557$ & $50 .001$ & $3 .693$ \\ \hline
 \multirow{5}{*}{$-\frac{7}{2}$} & \cellcolor{lightred} $\pi$ & \cellcolor{lightred} $3 .720$ &\cellcolor{lightred} $0 .553$ &\cellcolor{lightred} $4 .528$ &\cellcolor{lightred} $0 .0227$ & \cellcolor{lightred} $0 .332$ & \cellcolor{lightred} $0 .5$ &\cellcolor{lightred} $-0.5$ & \cellcolor{lightred}$0$ &\cellcolor{lightred} $8 .622$ & \cellcolor{lightred} $49 .918$  & \cellcolor{lightred} $1 .663$ \\ \cline{2-15}
 & $\frac{8 \pi }{9}$ & $3 .750$ & $0 .546$ & $12 .130$ & $0 .0241$ & $0 .331$ & $0 .470$ & $-0.542$ & $-0.117$ & $8 .663$ & $49 .871$ & $1 .955$ \\ \cline{2-15}
 & $\frac{10\pi }{9}$ & $3 .750$ & $0 .546$ & $11 .158$ & $0 .0241$ & $0 .331$ & $0 .530$ & $-0.458$ & $0 .117$ & $8 .663$ & $49 .870$ & $1 .955$ \\  \cline{2-15}
 & $\frac{9 \pi }{10}$ & $3 .744$ & $0 .547$ & $10 .125$ & $0 .0238$ & $0 .331$ & $0 .473$ & $-0.538$ & $-0.105$ & $8 .656$ & $49 .878$  & $1 .903$ \\ \cline{2-15}
 & $\frac{11\pi }{10}$ & $3 .745$ & $0 .547$ & $9 .248$ & $0 .0238$ & $0 .331$ & $0 .527$ & $-0.462$ & $0 .105$ & $8 .657$ & $49 .878$ & $1 .904$ \\ \hline
 \multirow{6}{*}{$-\frac{9}{2}$} & $\frac{2 \pi }{3}$ & $3 .724$ & $0 .329$ & $9 .712$ & $0 .0217$ & $0 .341$ & $0 .429$ & $-0.610$ & $-0.352$ & $8 .605$ & $49 .937$  & $3 .288$ \\ \cline{2-15}
 & $\frac{4\pi }{3}$ & $3 .724$ & $0 .329$ & $8 .224$ & $0 .0217$ & $0 .341$ & $0 .571$ & $-0.390$ & $0 .352$ & $8 .604$ & $49 .938$ & $3 .287$ \\ \cline{2-15}
 & $\frac{5 \pi }{8}$ & $3 .732$ & $0 .326$ & $15 .464$ & $0 .0239$ & $0 .340$ & $0 .424$ & $-0.610$ & $-0.396$ & $8 .566$ & $49 .990$ & $3 .454$ \\ \cline{2-15}
 & $\frac{11 \pi }{8}$ & $3 .731$ & $0 .326$ & $14 .061$ & $0 .0239$ & $0 .340$ & $0 .576$ & $-0.390$ & $0 .396$ & $8 .564$ & $49 .992$  & $3 .454$ \\ \cline{2-15}
 & $\frac{7 \pi }{10}$ & $3 .708$ & $0 .332$ & $17 .788$ & $0 .0199$ & $0 .342$ & $0 .433$ & $-0.607$ & $-0.317$ & $8 .610$ & $49 .931$ & $3 .147$ \\ \cline{2-15}
 & $\frac{13 \pi }{10}$ & $3 .708$ & $0 .332$ & $16 .242$ & $0 .0199$ & $0 .342$ & $0 .567$ & $-0.393$ & $0 .317$ & $8 .610$ & $49 .932$ & $3 .147$ \\ \hline \hline
\end{tabular}
\caption{\label{tab:bf}The best fit values of the lepton mixing angles, CP violation phases $\delta_{CP}$ and $\beta$, and the neutrino masses $m_2$ and $m_3$ for some typical values of $x$ and $\eta$ in new variants of Littlest seesaw model arising from the tri-direct CP approach with $S_4$. The predictions for the effective Majorana mass $m_{ee}$ are listed in the last column. We would like to remind that the lightest neutrino is massless $m_1=0$ for each case.}
\end{table}

We plot the best fit values of $\chi^2$ as a function of $\eta$ in figure~\ref{fig:correlation}, where five typical values of $x=-7/2, -4, -9/2, -5, -11/2$ are chosen for illustration. We notice that low $\chi^2$ values such as $\chi^2<10$ can be achieved. It is obvious that the values of $\chi^2$ is quite sensitive to the phase $\eta$, and the model can give very good fits to the leptonic mixing angles and the neutrino masses for certain values of $\eta$. We see that the experimental data can be described very well for $x=-7/2$ and $\eta$ around $\pi$. In table~\ref{tab:bf}, we show the best fit values of the mixing parameters and neutrino masses for some benchmark values of $x$ and $\eta$. Once the values of $x$ and $\eta$ are fixed, the light neutrino mass matrix $m_{\nu}$ would depend on only two free parameters $m_s$ and $m_a$ whose values can be determined by the neutrino mass squared differences $\Delta m^2_{21}$ and  $\Delta m^2_{31}$, then three lepton mixing angles and CP violation phases $\delta_{CP}$ and $\beta$ can be predicted. We see that the effective Majorana mass $m_{ee}$ is in the range of 1 and 3 meV such that it is impossible to be measured in foreseeable future. An particularly interesting example is the case of $x=-7/2$ and $\eta=\pi$, it predicts maximal atmospheric mixing angle $\theta_{23}=\pi/4$ and maximal Dirac phase $\delta_{CP}=-\pi/2$ which are favored by the present data from T2K and NO$\nu$A~\cite{Abe:2017uxa,Adamson:2017gxd}. The reason is because the general neutrino mass $m_\nu$ shown in Eq.~\eqref{eq:mnu} has a accidental $\mu\tau$ reflection symmetry in the case of $\eta=\pi$~\cite{King:2018kka}.

It is noteworthy that the cases predicting inverted neutrino masses can also be achieved from the tri-direct CP approach, although the neutrino mass spectrum is determined to be normal ordering in the present Littlest Seesaw variants, as shown in table~\ref{tab:bf}. In the following, we shall present an example giving inverted neutrino masses, for more other examples see~\cite{Ding:2018tuj}. Analogous to the above Littlest Seesaw variants, both lepton doublet $L$ and the atmospheric flavon $\phi_{\text{atm}}$ are assigned to $S_4$ triplet $\mathbf{3}$, the solar flavon $\phi_{\text{sol}}$ transforms as $\mathbf{3}'$ while the right-handed neutrino $N^c_{\mathrm{atm}}$ is $\mathbf{1}$ and $N^c_{\mathrm{sol}}$ is $\mathbf{1^\prime}$ of $S_4$. The residual flavor symmetry of the charged lepton sector is taken to be $G_{l}=Z^{T}_3$. The residual symmetries of the atmospheric neutrino and the solar neutrino sectors are $Z^{U}_{2}\times H^{\text{atm}}_{CP}$ and  $Z^{TU}_{2}\times H^{\text{sol}}_{CP}$ respectively with $H^{\text{atm}}_{CP}=\{1,U\}$ and $H^{\text{sol}}_{CP}=\{U,T\}$. Then it is easy to check that the alignments of $\phi_{\text{atm}}$ and $\phi_{\text{sol}}$ are $\langle\phi_{\text{atm}}\rangle\propto\left(0, 1, -1\right)^T$ and $\langle\phi_{\text{sol}}\rangle\propto\left(1,x\omega , x\omega^2 \right)^T$ respectively, where $x$ is a real parameter. As a consequence, the neutrino mass matrix is of the form
\begin{equation}
 m_{\nu}=m_{a}\begin{pmatrix}
 0 &~ 0 &~ 0 \\
 0 &~ 1 &~ -1 \\
 0 &~ -1 &~ 1 \\
\end{pmatrix}+m_{s}e^{i\eta}
\begin{pmatrix}
  1 &~ x \omega ^2 &~ x \omega  \\
 x \omega ^2 &~ x^2 \omega  &~ x^2 \\
 x \omega  &~ x^2 &~ x^2 \omega ^2 \\
\end{pmatrix}\,,
\end{equation}
where $m_{a}$ and $m_{s}$ are real parameters. We find the lepton mixing matrix is given by
\begin{equation}
\label{eq:PMNS_IO}\hskip-0.1in U_{PMNS}=\frac{1}{\sqrt{2}}
\begin{pmatrix}
 \frac{2 e^{-i \psi } \sin \theta }{\sqrt{2 +x^2}} &~ \frac{2 \cos \theta }{\sqrt{2 +x^2}} &~ -\frac{\sqrt{2} x}{\sqrt{2 +x^2}} \\
 -\cos \theta-\frac{x e^{-i \psi } \sin \theta }{\sqrt{2 +x^2}}  &~ e^{i \psi } \sin \theta-\frac{x \cos \theta }{\sqrt{2 +x^2}}  &~ -\frac{\sqrt{2} }{\sqrt{2 +x^2}} \\
 \cos \theta -\frac{x e^{-i \psi } \sin \theta }{\sqrt{2 +x^2}} &~ -e^{i \psi } \sin \theta -\frac{x \cos \theta }{\sqrt{2 +x^2}} &~ -\frac{\sqrt{2}}{\sqrt{2 +x^2}} \\
\end{pmatrix}P_{\nu}\,,
\end{equation}
with
\begin{equation}
P_{\nu}=\text{diag}(e^{i(\psi+\rho)/2}, e^{i(-\psi+\sigma)/2},1)\,.
\end{equation}
The three lepton mixing angles can be read off as,
\begin{equation}
 \sin^2\theta_{13}=\frac{x^2}{2+x^2}\,, \quad
  \sin^2\theta_{12}=\cos ^2\theta \,, \quad
\sin^2\theta_{23}=\frac{1}{2}\,.
\end{equation}
It is notable that the atmospheric mixing angle $\theta_{23}$ is exactly maximal. Moreover, the two CP rephasing invariants are given by
\begin{equation}
J_{CP}=-\frac{x \sin 2\theta  \sin \psi }{2\left(2 +x^2\right)^{3/2}}\,, \qquad I_{1}=-\frac{ \sin ^22 \theta \sin (\rho -\sigma )}{\left(2 +x^2\right)^2}\,.
\end{equation}
The experimentally measured values of lepton mixing angles and neutrino masses can be accommodated well in this case, e.g.
\begin{eqnarray}
\nonumber && x=-0.213, \quad  \eta=-0.0171\pi, \quad
m_{a}=25.670\,\text{meV},\quad r=1.823\,, \\
\nonumber && \sin^2\theta_{13}=0.0223, \quad  \sin^2\theta_{12}=0.307, \quad \sin^2\theta_{23}=0.5, \quad \delta_{CP}/\pi=0.975, \quad \beta/\pi=-0.174, \\
&&  m_1=49.193\,\text{meV}, \quad m_2=49.940\,\text{meV},  \quad m_3=0, \quad m_{ee}=46.579\,\text{meV}\,.
\end{eqnarray}
From the examples of Littlest Seesaw model and its variants studied above, we see that the light neutrino mass matrix $m_{\nu}$ is generally predicted to depend on four parameters $m_a$, $m_s$, $x$ and $\eta$ in the tri-direct CP approach, and $m_{\nu}$ would depend on only two parameters $m_a$ and $m_s$ once $x$ and $\eta$ are fixed to certain simple regular values by explicit superpotential alignment terms in a concrete model. Nevertheless, the neutrino mass matrix $m_{\nu}$ generally involve six complex parameters in the original two right-handed neutrino model. Obviously the tri-direct CP model is rather predictive beyond all doubt.

\section{\label{sec:extension}Extending the tri-direct CP approach to three right-handed neutrino models  }

Motivated by the principle of minimality and the idea of constrained sequential dominance, in section~\ref{sec:framework} we have assumed that there are only two right-handed neutrinos and the third one is approximately decoupled. In fact, the tri-direct CP approach is a general paradigm of neutrino mass model building, and it is not mandatory to have two right-handed neutrinos. This approach can be straightforwardly extended to the conventional seesaw model with three right-handed neutrinos denote as $N^c_{\mathrm{atm}}$, $N^c_{\mathrm{sol}}$ and $N^c_{\mathrm{dec}}$. Then the Lagrangian for the charged lepton and neutrino masses takes the form
\begin{eqnarray}
\nonumber\mathcal{L}&=&-y_{l}L\phi_{l}E^{c}-y_{\mathrm{atm}}L\phi_{\mathrm{atm}}N^c_{\mathrm{atm}}-y_{\mathrm{sol}}L\phi_{\mathrm{sol}}N^c_{\mathrm{sol}}-y_{\mathrm{dec}}L\phi_{\mathrm{dec}}N^c_{\mathrm{dec}}
-\frac{1}{2}x_{\mathrm{atm}}\xi_{\mathrm{atm}}N^c_{\mathrm{atm}}N^c_{\mathrm{atm}}\\
\label{eq:Lagrangian_extens}&&-\frac{1}{2}x_{\mathrm{sol}}\xi_{\mathrm{sol}}N^{c}_{\mathrm{sol}}N^c_{\mathrm{sol}}-\frac{1}{2}x_{\mathrm{dec}}\xi_{\mathrm{dec}}N^{c}_{\mathrm{sol}}N^c_{\mathrm{dec}}
+\text{h.c.}\,,
\end{eqnarray}
where two additional terms related to $N^c_{\mathrm{dec}}$ appear in comparison with Eq.~\eqref{eq:Lagrangian}. The field $\phi_{\mathrm{dec}}$ can be either Higgs fields or combination of the electroweak Higgs doublet together with flavons, and it transform as triplet under the flavor symmetry $G_{f}$. The Majoron flavon $\xi_{\mathrm{dec}}$ are standard model and $G_f$ singlet. Similar to section~\ref{sec:framework}, we assume that the residual subgroups $G_{l}$, $G_{\text{atm}}\rtimes H^{\text{atm}}_{CP}$, $G_{\text{sol}}\rtimes H^{\text{sol}}_{CP}$ and $G_{\text{dec}}\rtimes H^{\text{dec}}_{CP}$ are preserved by the Yukawa interaction terms of the charged leptons $E^c$, the atmospheric neutrino $N^c_{\mathrm{atm}}$, the solar neutrino $N^c_{\mathrm{sol}}$ and the decoupled neutrino $N^c_{\mathrm{dec}}$ respectively. The vacuum alignments $\langle\phi_{\rm atm}\rangle$, $\langle\phi_{\rm sol}\rangle$ and $\langle\phi_{\rm dec}\rangle$ are dictated by the residual flavor and CP symmetries, and they constitute three columns of the Dirac neutrino mass matrix,
\begin{eqnarray}
\nonumber m_{D}&=&\begin{pmatrix}
y_{\text{atm}}\langle\phi_{\text{atm}}\rangle,  ~&~ y_{\text{sol}}\langle\phi_{\text{sol}}\rangle, ~&~ y_{\text{dec}}\langle\phi_{\rm dec}\rangle
\end{pmatrix},\\[0.1in]
m_{N}&=&\begin{pmatrix}
x_{\textrm{atm}}\langle\xi_{\text{atm}}\rangle ~&~ 0 ~&~ 0 \\
0  ~& ~ x_{\textrm{sol}}\langle\xi_{\text{sol}}\rangle ~&~ 0 \\
0 ~&~  0 ~&~ x_{\textrm{dec}}\langle\xi_{\text{dec}}\rangle
\end{pmatrix}\,.
\end{eqnarray}
The light neutrino mass matrix given by the seesaw formula is
\begin{equation}\label{eq:mnu1}
m_{\nu}=-\frac{y^2_{\text{atm}}}{x_{\text{atm}}}\frac{\langle\phi_{\rm atm}\rangle\langle\phi_{\rm atm}\rangle^T}{\langle\xi_{\text{atm}}\rangle}
-\frac{y^2_{\text{sol}}}{x_{\text{sol}}}\frac{\langle\phi_{\rm sol}\rangle\langle\phi_{\rm sol}\rangle^T}{\langle\xi_{\text{sol}}\rangle}
-\frac{y^2_{\text{dec}}}{x_{\text{dec}}}\frac{\langle\phi_{\rm dec}\rangle\langle\phi_{\rm dec}\rangle^T}{\langle\xi_{\text{dec}}\rangle}\,.
\end{equation}
In the limit $\langle\xi_{\text{dec}}\rangle\gg \langle\xi_{\text{atm}}\rangle, \langle\xi_{\text{sol}}\rangle$, it reduces to the setup discussed in section~\ref{sec:framework}. Here we shall give an example for illustration. The lepton doublet $L$, the atmospheric flavon $\phi_{\text{atm}}$ and the flavon $\phi_{\text{dec}}$ are assumed to transforms as $\mathbf{3}$ under $S_4$, the solar flavon $\phi_{\text{sol}}$ transforms as $\mathbf{3}'$ while the right-handed neutrinos $N^c_{\mathrm{atm}}$ and $N^c_{\mathrm{dec}}$ are $S_4$ singlet $\mathbf{1}$ and $N^c_{\mathrm{sol}}$ is $\mathbf{1^\prime}$ under $S_4$. The flavor group $S_4$ and CP symmetry are broken to $G_{l}=Z^{T}_3$ in the charged lepton sector. The residual symmetries of the atmospheric neutrino, the solar neutrino and the decoupled neutrino sectors are $Z^{U}_{2}\times H^{\text{atm}}_{CP}$, $Z^{SU}_{2}\times H^{\text{sol}}_{CP}$ and $Z^{TST^2}_{2}\times H^{\text{dec}}_{CP}$ respectively, where the residual CP symmetries are given by $H^{\text{atm}}_{CP}=\{1,U\}$, $H^{\text{sol}}_{CP}=\{1,SU\}$ and $H^{\text{dec}}_{CP}=\{SU,T^2STU\}$. The vacuum alignments of $\phi_{\text{atm}}$, $\phi_{\text{sol}}$ and $\phi_{\text{dec}}$ are constrained by the residual symmetry to take the following form
\begin{equation}
\langle\phi_{\text{atm}}\rangle\propto\left(0, 1, -1\right)^T, \quad \langle\phi_{\text{dec}}\rangle\propto\left(1, \omega^2, \omega\right)^T, \quad\langle\phi_{\text{sol}}\rangle\propto\left(1,3 ,-1 \right)^T\,.
\end{equation}
Then we can read out the light neutrino mass matrix
\begin{equation}
m_{\nu}=m_{a}\begin{pmatrix}
0 &~ 0 &~ 0 \\
0 &~ 1 &~ -1 \\
0 &~ -1 &~ 1 \\
\end{pmatrix}+m_{a}r_{1}e^{i\eta_{1}}
\begin{pmatrix}
 1 & -1 & 3 \\
 -1 & 1 & -3 \\
 3 & -3 & 9 \\
\end{pmatrix}+m_{a}r_{2}e^{i\eta_{2}}
\begin{pmatrix}
 1 & \omega  & \omega ^2 \\
 \omega  & \omega ^2 & 1 \\
 \omega ^2 & 1 & \omega  \\
\end{pmatrix}\,,
\end{equation}
where $m_{a}$, $r_{1}$, $r_{2}$, $\eta_{1}$ and $\eta_{2}$ are real.
Excellent agreement with experimental data can be achieved in this scenario, and a numerical benchmark is
\begin{eqnarray}
\nonumber &&\hskip-0.15in  m_{a}=26.707\,\text{meV}, \quad  r_{1}=0.101, \quad r_{2}=0.00300, \quad\eta_{1}=2\pi/3,\quad \eta_2=0\,,\\
\nonumber && \hskip-0.15in  \sin^2\theta_{13}=0.0221,\quad  \sin^2\theta_{12}=0.318, \quad \sin^2\theta_{23}=0.486\,,\\
\nonumber && \hskip-0.15in   \delta_{CP}=-0.519\pi, \quad \alpha_{21}=-0.533\pi,\quad \alpha_{31}=-0.135\pi \,, \\
&& \hskip-0.15in     m_1=0.120\,\text{meV}, \quad m_2=8.601\,\text{meV},  \quad m_3=49.942\,\text{meV}, \quad m_{ee}=2.648\,\text{meV}\,.
\end{eqnarray}
It is interesting to perform a comprehensive study of lepton mixing patterns which can be obtained from $S_4$ and other flavor groups from the tri-direct CP approach with three right-handed neutrinos. The phenomenological implications and model building aspects of such a scheme also deserve further investigation. These topics goes far beyond the scope of this paper since they deserve a dedicated full work of its own.

Before closing this section, we would like to compare the tri-direct CP approach with other flavor symmetry model building schemes such as direct approach, semidirect approach and indirect approach.  In common with the indirect approach~\cite{King:2013eh}, each column of the Dirac neutrino mass matrix is effectively promoted to a flavon field which transforms as a triplet under the flavor symmetry $G_{f}$. The tri-direct CP approach generalizes the indirect approach to constrain the vacuum alignment by residual flavor and CP symmetries. As regards the direct~\cite{King:2013eh} and semidirect models~\cite{Feruglio:2012cw,Ding:2013hpa}, a residual symmetry $Z_2\times Z_2$ or $Z_2\times CP$ is preserved by the neutrino mass matrix, the lepton mixing matrix is fully determined by residual symmetry or determined in terms of a single real parameter $\theta$ while neutrino masses are not constrained~\cite{King:2013eh,Feruglio:2012cw,Ding:2013hpa}. In direct and semidirect models, only the structure of
flavor symmetry group and the remnant symmetries are assumed, the neutrino mass generation mechanism and the breaking mechanism of flavor and CP symmetries are irrelevant such that one can analyze the predictions for lepton mixing parameters in a model-independent way~\cite{King:2013eh,Feruglio:2012cw,Ding:2013hpa}. Similar to the
semidirect models with residual symmetry $Z_2\times CP$, the assumed residual symmetry of tri-direct CP approach fixes one column of the neutrino mixing matrix to be $\langle\phi_{\text{atm}}\rangle\times\langle\phi_{\text{sol}}\rangle$ which could depend on a real parameter $x$. However, there is no common residual symmetry of the light neutrino mass matrix in the tri-direc CP approach since the residual symmetries $G_{\text{atm}}\rtimes H^{\text{atm}}_{CP}$ and $G_{\text{sol}}\rtimes H^{\text{sol}}_{CP}$ of the atmospheric and solar neutrino sectors are generally different. In contrast with direct and semidirect approaches, the tri-direct CP approach can predict neutrino masses. As shown in sections~\ref{LSStri-direct} and~\ref{sec:new_LSS}, the light neutrino mass matrix $m_{\nu}$ generally depends on only four real parameters $m_a$, $m_s$, $x$ and $\eta$ such that lepton mixing angles, CP phases and neutrino masses are strongly correlated with each other in tri-direct CP models. We remark that the tri-direct CP approach assumes type-I seesaw mechanism for neutrino mass generation, consequently  it is not applicable in radiative neutrino mass models and so on.

\section{\label{sec:model} A concrete model  }

In this section, we shall construct an explicit model based on the model independent analysis in section~\ref{sec:new_LSS}. The flavor symmetry $S_4$ together with CP symmetry is imposed in the model. The auxiliary symmetry is taken to be $Z_{5}\times Z_6 \times Z_8$ to ensure the needed vacuum alignment and to forbid unwanted couplings. The auxiliary symmetry $Z_{6}$ is helpful to reproduce the observed charged lepton mass hierarchies, and it imposes different powers of flavon fields for the electron, muon and tauon terms. The shaping symmetry $Z_8$ disentangles the charged lepton sector from the neutrino sector, and $Z_5$ further distinguishes the atmospheric neutrino sector from the solar neutrino sector. Moreover, it is straightforward to show that such a symmetry is sufficient to suppress higher dimensional terms. The spontaneous breaking of $S_4$ and CP symmetries to the residual symmetries $Z^T_3$, $Z^{TST^2}_2\times X_{\text{atm}}$ and $Z_2^U\times X_{\text{sol}}$ in the charged lepton, atmospheric neutrino and solar neutrino sectors are achieved in the model, where the residual CP transformations $X_{\text{atm}}=SU$ and $X_{\text{sol}}=U$. As a consequence, the desired vacuum configurations in Eqs.~(\ref{eq:VEV_atm}, \ref{eq:VEV_sol}) are naturally produced. The solar alignment parameter is fixed to be $x=-7/2$ with $\eta=\pi$ through the dynamical terms in the potential. We formulate our model in the framework of supersymmetry since the minimization of the scalar potential would be considerably simplified. The three generations of left-handed lepton doublets $L$ are embedded into a triplet $\mathbf{3}$, while the right-handed charged leptons $e^{c}$, $\mu^{c}$ and $\tau^{c}$ all transform as $\mathbf{1}$ yet they carry different charges of shaping symmetry. The two right-handed neutrinos $\nu^{c}_{\text{atm}}$ and $\nu^{c}_{\text{sol}}$ are assigned to $\mathbf{1}$ and $\mathbf{1}'$ of $S_4$ respectively. The $S_4$ flavor symmetry is then broken by suitable flavons which are singlets under the standard model gauge group. The fields of the model and their classification under the symmetry are summarized in table~\ref{tab:field}. Similar to other flavor models in the literature~\cite{Altarelli:2010gt,Ishimori:2010au,King:2013eh,King:2014nza,King:2015aea,King:2017guk}, additional flavon fields besides the necessary flavons $\phi_{a}$, $\phi_{s}$, $\xi_{a}$ and $\xi_{s}$ are required in order to achieve the desired vacuum configuration. Their transformation properties are shown in table~\ref{tab:field}. We would like to remind the readers that we adopt the convention of~\cite{Ding:2013hpa} for the $S_4$ group, and all the Clebsch-Gordan coefficients have been listed in the Appendix of ~\cite{Ding:2013hpa}.

\begin{table}[t!]
\renewcommand{\tabcolsep}{0.4mm}
\begin{center}
\begin{tabular}{|c|c|c|c|c|c|c|c||c|c||c|c||c|c|c|c|c|c|c||c|c||c|c||c|c|c|c|c|c|c|}\hline\hline
 & $L$ & $e^c$  &  $\mu^c$ &  $\tau^c$  & $\nu^{c}_{\text{atm}}$ & $\nu^{c}_{\text{sol}}$  &$H_{u,d}$ & $\eta_{l}$ &  $\phi_{l}$   & $\xi_{a}$ & $\phi_{a}$  & $\xi_{s}$ & $\zeta_{s}$ & $\eta_{s}$ & $\chi_{s}$ & $\psi_{s}$ &  $\varphi_{s}$ &  $\phi_{s}$  &  $\xi^0_{l}$  &  $\phi^0_{l}$ & $\xi^0_{a}$ & $\phi^0_{a}$ & $\kappa^0$  & $\rho^0$ & $\sigma^0$ & $\eta^0$ & $\chi^0$ & $\phi^0_{s}$  & $\xi^0_{s}$ \\ \hline

$S_4$  & $\mathbf{3}$ &  $\mathbf{1}$  &  $\mathbf{1}$ & $\mathbf{1}$  & $\mathbf{1}$ & $\mathbf{1^\prime}$ & $\mathbf{1}$  &  $\mathbf{2}$ & $\mathbf{3}$  &  $\mathbf{1}$ & $\mathbf{3}$  &  $\mathbf{1}$ &  $\mathbf{1}$  & $\mathbf{2}$  &  $\mathbf{3}^{\prime}$ & $\mathbf{3}^{\prime}$  & $\mathbf{3}^{\prime}$  &  $\mathbf{3}^{\prime}$  &  $\mathbf{1}$ &  $\mathbf{3}^{\prime}$ & $\mathbf{1}$  &  $\mathbf{3^\prime}$ & $\mathbf{1}$  &  $\mathbf{2}$ & $\mathbf{2}$  &  $\mathbf{2}$ & $\mathbf{3}^{\prime}$ &  $\mathbf{3}^{\prime}$ &  $\mathbf{1}$ \\

$Z_5$ & $1$ & $1$  & $1$ & $1$ & $1$  &  $\omega^4_5$  & $1$  &  $1$ & $1$ & $1$ & $1$  & $\omega^2_5$  & $\omega_5$ & $\omega_5$ & $\omega^2_5$  & $\omega^4_5$ & $\omega^3_5$ & $\omega_5$  & $1$ &   $1$ & $1$ &   $1$  & $\omega^4_5$    & $\omega_5$ & $\omega^2_5$ & $\omega^3_5$  & $\omega_5$ & $\omega^3_5$  & $\omega^3_5$ \\

$Z_6$ & $1$ &  $\omega^3_6$  &  $\omega^4_6$  & $\omega^5_6$  & $\omega_6$ & $1$ & $1$ & $\omega_6$ &  $\omega_6$ & $\omega^4_6$ &  $\omega^5_6$ &  $1$ &  $1$ &  $1$  & $1$ & $1$ &   $1$ &  $1$  & $\omega^4_6$ &  $\omega^4_6$ & $\omega^2_6$ &  $\omega^2_6$  &  $1$ &  $1$ &  $1$ &   $1$ &  $1$ &  $1$ &  $1$  \\

$Z_8$ & $1$  & $1$   & $1$   & $1$  & $\omega_8$  & $\omega^2_8$  & $1$  & $1$  &  $1$  &  $\omega^6_8$  & $\omega^7_8$  & $\omega^4_8$ & $\omega^4_8$ & $\omega_8$  &   $\omega_8$   & $\omega_8$   & $\omega_8$ &  $\omega^6_8$  &  $1$   & $1$ & $\omega^2_8$   & $\omega^2_8$  &  $\omega^6_8$   & $\omega^6_8$ & $\omega^6_8$   & $\omega^6_8$  &  $\omega^6_8$  &  $\omega^6_8$ &  $\omega^4_8$ \\ \hline\hline

\end{tabular}
\caption{\label{tab:field}Fields and their transformation properties under the flavor symmetry $S_4\times Z_{5}\times Z_6 \times Z_8$, where the phases are $\omega_5=e^{2\pi i/5}$, $\omega_6=e^{\pi i/3}$ and $\omega_8=e^{\pi i/4}$. }
\end{center}
\end{table}

\subsection{\label{subsec:alignment}Vacuum alignment}

We will use the supersymmetric $F-$term alignment mechanism to generate the flavon VEVs in Eqs.~\eqref{eq:VEV_atm} and \eqref{eq:VEV_sol}. A $U(1)_R$ symmetry related to $R-$parity and the presence of driving fields in the flavon superpotential are common features of this mechanism. The driving fields indicated with the superscript ``0'' and the symmetry assignments are collected in table~\ref{tab:field}. As usual, the VEVs of the driving fields are assumed to be vanishing. In the supersymmetric limit, the $F-$terms of the driving fields have to vanish such that the vacuum of the flavons gets aligned. The minimization equation of the scalar potential for the charged lepton, atmospheric neutrino and solar neutrino sectors are separated from each other at the renormalizable level.

At leading order, the most general driving superpotential $w_d$ invariant under $S_4\times Z_{5}\times Z_6 \times Z_8$ is given by
\begin{equation}
w_{d}=w_{d}^l+w_{d}^{\text{atm}}+w_{d}^{\text{sol}}\,,
\end{equation}
where the three parts $w_{d}^l$, $w_{d}^{\text{atm}}$ and $w_{d}^{\text{sol}}$ read
\begin{eqnarray}
\nonumber&&\hskip-0.15in w_{d}^l=g_{1}\xi^0_{l}\left(\eta_{l}\eta_{l}\right)_{\mathbf{1}}+g_2\xi^0_{l}\left(\phi_{l}\phi_{l}\right)_{\mathbf{1}}
+g_3\left(\phi^{0}_{l}\left(\eta_{l}\phi_{l}\right)_{\mathbf{3^\prime}}\right)_{\mathbf{1}}
+g_4\left(\phi^{0}_{l}\left(\phi_{l}\phi_{l}\right)_{\mathbf{3^\prime}}\right)_{\mathbf{1}}\,,\\
\nonumber&&\hskip-0.15in w_{d}^{\text{atm}}=M_{\xi_{a}}\xi^0_{a}\xi_{a}
+h_1\xi^0_{a}\left(\phi_{a}\phi_{a}\right)_{\mathbf{1}}
+h_{2}\left(\phi^{0}_{a}\left(\phi_{a}\phi_{a}\right)_{\mathbf{3^\prime}}\right)_{\mathbf{1}}\,,\\
\nonumber&&\hskip-0.15in w_{d}^{\text{sol}}=f_{1}\kappa^0(\chi_{s}\psi_{s})_{\mathbf{1}}
+f_{2}\kappa^0(\varphi_{s}\varphi_{s})_{\mathbf{1}}+f_{3}\left(\rho^0\left(\chi_{s}\chi_{s}\right)_{\mathbf{2}}\right)_{\mathbf{1}}
+f_{4}\left(\sigma^0\left(\psi_{s}\psi_{s}\right)_{\mathbf{2}}\right)_{\mathbf{1}}
+f_{5}\left(\eta^0\left(\eta_{s}\eta_{s}\right)_{\mathbf{2}}\right)_{\mathbf{1}} \\
\nonumber &&\qquad+f_{6}\left(\eta^0\left(\varphi_{s}\psi_{s}\right)_{\mathbf{2}}\right)_{\mathbf{1}}
+f_{7}\left(\chi^0\left(\chi_{s}\chi_{s}\right)_{\mathbf{3^\prime}}\right)_{\mathbf{1}}
 +f_{8}\left(\chi^0\left(\eta_{s}\varphi_{s}\right)_{\mathbf{3^\prime}}\right)_{\mathbf{1}}
+f_{9}\left(\phi^0_{s}\left(\psi_{s}\varphi_{s}\right)_{\mathbf{3^\prime}}\right)_{\mathbf{1}} \\
\label{eq:wd}&&\qquad +f_{10}\zeta_{s}\left(\phi^0_{s}\phi_{s}\right)_{\mathbf{1}}
+M_{\xi_{s}}\xi^0_{s}\xi_{s}+f_{11}\xi^0_{s}\left(\phi_{s}\phi_{s}\right)_{\mathbf{1}}\,,
\end{eqnarray}
where $(\ldots)_{\mathbf{r}}$ refers to a contraction of the $S_4$ indices into the representation $\mathbf{r}$. All the coupling constants are real parameters since the theory is required to be invariant under the generalized CP transformations. The driving superpotential $w^{l}_d$ is responsible for the alignment of $\eta_{l}$ and $\phi_{l}$. The equations for the vanishing of the derivatives of $w^{l}_d$ with respect to each component of the driving fields $\xi^{0}_{l}$ and $\phi^{0}_{l}$ are
\begin{eqnarray}
\label{eq:align_ch}
\nonumber&&\frac{\partial w_{d}^l}{\partial\xi^{0}_{l}}=2 g_{1} \eta_{l_{1}} \eta_{l_{2}}+g_{2} \left(\phi_{l_{1}}^2+2 \phi_{l_{2}} \phi_{l_{3}}\right)=0\,,\\
\nonumber&&\frac{\partial w_{d}^l}{\partial\phi^{0}_{l_{1}}}=g_{3} (\eta_{l_{1}} \phi_{l_{2}}-\eta_{l_{2}} \phi_{l_{3}})+2g_{4} \left( \phi_{l_{1}}^2- \phi_{l_{2}} \phi_{l_{3}}\right)=0\,,\\
\nonumber&&\frac{\partial w_{d}^l}{\partial\phi^{0}_{l_{2}}}=g_{3} (\eta_{l_{1}} \phi_{l_{1}}-\eta_{l_{2}} \phi_{l_{2}})+2g_{4} \left( \phi_{l_{2}}^2- \phi_{l_{1}} \phi_{l_{3}}\right)=0\,,\\
&&\frac{\partial w_{d}^l}{\partial\phi^{0}_{l_{3}}}=g_{3} (\eta_{l_{1}} \phi_{l_{3}}-\eta_{l_{2}} \phi_{l_{1}})+2g_{4} \left( \phi_{l_{3}}^2- \phi_{l_{1}} \phi_{l_{2}}\right)=0\,.
\end{eqnarray}
These equations are satisfied by the alignment
\begin{equation}
\label{eq:vacuum_ch}\langle\eta_{l}\rangle=\left(0, v_{\eta_{l}}\right)^T,\qquad \langle\phi_{l}\rangle=\left(0, v_{\phi_{l}},0\right)^T, \quad \text{with} \quad v_{\phi_{l}}=\frac{g_{3}}{2g_{4}}v_{\eta_{l}}\,,
\end{equation}
where $v_{\eta_{l}}$ is undetermined. In the atmospheric neutrino sector, the $F-$flatness condition of the driving fields $\xi^0_{a}$ and $\phi^{0}_{a}$ leads to
\begin{eqnarray}
\nonumber&&\frac{\partial w_{d}^{\text{atm}}}{\partial\xi^0_{a}}=M_{\xi_{a}} \xi_{a}+h_{1} \left(\phi_{a_{1}}^2+2 \phi_{a_{2}} \phi_{a_{3}}\right)=0\,, \\
\nonumber&&\frac{\partial w_{d}^{\text{atm}}}{\partial\phi^{0}_{a_{1}}}=h_{2} \left(2 \phi_{a_{1}}^2-2 \phi_{a_{2}} \phi_{a_{3}}\right)=0\,,\\
\nonumber&&\frac{\partial w_{d}^{\text{atm}}}{\partial\phi^{0}_{a_{2}}}=h_{2} \left(2 \phi_{a_{2}}^2-2 \phi_{a_{1}} \phi_{a_{3}}\right)=0\,,\\
&&\frac{\partial w_{d}^{\text{atm}}}{\partial\phi^{0}_{a_{3}}}=h_{2} \left(2 \phi_{a_{3}}^2-2 \phi_{a_{1}} \phi_{a_{2}}\right)=0\,,
\end{eqnarray}
from which we can extract the vacuum expectation values of $\xi_{a}$ and $\phi_{a}$\footnote{The alignment of $\phi_a$ can also be along the direction $(1, 1, 1)^{T}$ and it is related to the chosen one in Eq.~\eqref{eq:vacuum_atm} by a $T$ transformation.},
\begin{equation}
\label{eq:vacuum_atm}\langle\xi_{a}\rangle=v_{\xi_{a}},\quad \langle\phi_{a}\rangle=v_{\phi_{a}}\left(1, \omega^2, \omega\right)^T,  \quad \text{with} \quad
v^2_{\phi_{a}} =-\frac{M_{\xi_{a}}}{3h_1}v_{\xi_{a}}\,.
\end{equation}
We see that the desired atmospheric flavon alignment of $\phi_{a}$ in Eq.~\eqref{eq:VEV_atm} is realized, and it preserves the subgroup $Z^{TST^2}_2$. Moreover, the ratio $v^2_{\phi_{a}}/v_{\xi_{a}}=-\frac{M_{\xi_{a}}}{3h_1}$
contributing to the parameter $m_a$ is determined to be real. Then we proceed to derive the solar flavon alignment in a short sequence of steps. The $F-$term conditions of the driving field $\rho^{0}$ and $\sigma^0$ read
\begin{equation}
\begin{aligned}
&\frac{\partial w_{d}^{\text{sol}}}{\partial\rho^{0}_1}=f_{3} \left(2 \chi_{s_{1}} \chi_{s_{2}}+\chi_{s_{3}}^2\right)=0\,,\\
&\frac{\partial w_{d}^{\text{sol}}}{\partial\rho^{0}_2}=f_{3} \left(2 \chi_{s_{1}} \chi_{s_{3}}+\chi_{s_{2}}^2\right)=0\,,\\
&\frac{\partial w_{d}^{\text{sol}}}{\partial\sigma^{0}_1}=f_{4} \left(2 \psi_{s_{1}} \psi_{s_{2}}+\psi_{s_{3}}^2\right)=0\,,\\
&\frac{\partial w_{d}^{\text{sol}}}{\partial\sigma^{0}_2}=f_{4} \left(2 \psi_{s_{1}} \psi_{s_{3}}+\psi_{s_{2}}^2\right)=0\,.\\
\end{aligned}
\end{equation}
A solution to this set of equations is given by
\begin{equation}\label{eq:vacuum_solar_1}
\langle\chi_{s}\rangle=v_{\chi_{s}}(1,0,0)^T, \qquad
\langle\psi_{s}\rangle=v_{\psi_{s}}\left(1, -2,  -2\right)^T \,.
\end{equation}
Subsequently the $F-$flatness of the driving field $\eta^{0}$ and $\chi^0$ leads to
\begin{eqnarray}
\nonumber&&\frac{\partial w_{d}^{\text{sol}}}{\partial\eta^{0}_1}=f_{5} \eta_{s_{1}}^2+f_{6} (\varphi_{s_{1}} \psi_{s_{2}}+\varphi_{s_{2}} \psi_{s_{1}}+\varphi_{s_{3}} \psi_{s_{3}})=0\,,\\
\nonumber&&\frac{\partial w_{d}^{\text{sol}}}{\partial\eta^{0}_2}=f_{5} \eta_{s_{2}}^2+f_{6} (\varphi_{s_{1}} \psi_{s_{3}}+\varphi_{s_{2}} \psi_{s_{2}}+\varphi_{s_{3}} \psi_{s_{1}})=0\,,\\
\nonumber&&\frac{\partial w_{d}^{\text{sol}}}{\partial\chi^{0}_1}=2f_{7} \left( \chi_{s_{1}}^2- \chi_{s_{2}} \chi_{s_{3}}\right)+f_{8} (\eta_{s_{1}} \varphi_{s_{2}}+\eta_{s_{2}} \varphi_{s_{3}})=0\,,\\
\nonumber&&\frac{\partial w_{d}^{\text{sol}}}{\partial\chi^{0}_2}=2f_{7}\left( \chi_{s_{2}}^2- \chi_{s_{1}} \chi_{s_{3}}\right)+f_{8} (\eta_{s_{1}} \varphi_{s_{1}}+\eta_{s_{2}} \varphi_{s_{2}})=0\,,\\
&&\frac{\partial w_{d}^{\text{sol}}}{\partial\chi^{0}_3}=2f_{7} \left( \chi_{s_{3}}^2- \chi_{s_{1}} \chi_{s_{2}}\right)+f_{8} (\eta_{s_{1}} \varphi_{s_{3}}+\eta_{s_{2}} \varphi_{s_{1}})=0\,,
\end{eqnarray}
which generate the alignment
\begin{equation}\label{eq:vacuum_solar_2}
\langle\eta_{s}\rangle=v_{\eta_{s}}\left(1,1\right)^T,\qquad
\langle\varphi_{s}\rangle=v_{\varphi_{s}}\left(1, -1,  -1\right)^T \,,
\end{equation}
with
\begin{equation}
\label{eq:solar_VEV_1}v_{\varphi_{s}}=\frac{f_{5} v^2_{\eta_{s}}}{f_6 v_{\psi_{s}}},\qquad v^2_{\chi_{s}}=\frac{f_{5}f_{8} v^3_{\eta_{s}}}{f_{6}f_{7} v_{\psi_{s}}}\,.
\end{equation}
Furthermore, we notice that the contraction of $\langle\psi_s\rangle$ and $\langle\varphi_s\rangle$ to an $S_4$ triplet $\mathbf{3}'$ is of the form
\begin{equation}
\left(\langle\psi_s\rangle\langle\varphi_s\rangle\right)_{\mathbf{3}'}\propto\begin{pmatrix}
-2\\
7\\
7
\end{pmatrix}\,.
\end{equation}
Here we have used the $S_4$ contraction rule for $\mathbf{3}'\otimes\mathbf{3}'\rightarrow\mathbf{3}'$: $(\alpha\beta)_{\mathbf{3}'}=( 2\alpha_1\beta_1-\alpha_2\beta_3-\alpha_3\beta_2,  2\alpha_3\beta_3-\alpha_1\beta_2-\alpha_2\beta_1,                     2\alpha_2\beta_2-\alpha_1\beta_3-\alpha_3\beta_1)^{T}$, where $\alpha=(\alpha_1, \alpha_2, \alpha_3)^{T}$ and $\beta=(\beta_1, \beta_2, \beta_3)^{T}$ transform as $\mathbf{3}'$~\cite{Ding:2013hpa}. Therefore the solar flavon alignment arises from the $\phi^0_{s}$ dependent driving terms $f_{9}\left(\phi^0_{s}\left(\psi_{s}\varphi_{s}\right)_{\mathbf{3^\prime}}\right)_{\mathbf{1}}+f_{10}\zeta_{s}\left(\phi^0_{s}\phi_{s}\right)_{\mathbf{1}}$, and accordingly the minimization equations of the scalar potential are given by
\begin{eqnarray}
\nonumber&&\frac{\partial w_{d}^{\text{sol}}}{\partial\phi^{0}_{s_{1}}}=f_{10}\zeta_{s}\phi_{s_{1}}+f_{9} (2 \varphi_{s_{1}} \psi_{s_{1}}-\varphi_{s_{2}} \psi_{s_{3}}-\varphi_{s_{3}} \psi_{s_{2}})=0\,,\\
\nonumber&&\frac{\partial w_{d}^{\text{sol}}}{\partial\phi^{0}_{s_{2}}}=f_{10}\zeta_{s}\phi_{s_{3}}+f_{9} (-\varphi_{s_{1}} \psi_{s_{3}}+2 \varphi_{s_{2}} \psi_{s_{2}}-\varphi_{s_{3}} \psi_{s_{1}})=0\,,\\
&&\frac{\partial w_{d}^{\text{sol}}}{\partial\phi^{0}_{s_{3}}}=f_{10}\zeta_{s}\phi_{s_{2}}+f_{9} (-\varphi_{s_{1}} \psi_{s_{2}}-\varphi_{s_{2}} \psi_{s_{1}}+2 \varphi_{s_{3}} \psi_{s_{3}})=0\,,
\end{eqnarray}
which uniquely determine the solar alignment,
\begin{equation}
\label{eq:vacuum_solar_3}
\langle\zeta_{s}\rangle=v_{\zeta_{s}},\quad \langle\phi_{s}\rangle=v_{\phi_{s}}\left(1, -7/2,  -7/2\right)^T \,, \quad \text{with} \quad
v_{\phi_{s}}=\frac{2 f_{9} v_{\varphi_{s}}v_{\psi_{s}}}{f_{10}v_{\zeta_{s}}}\,.
\end{equation}
Finally the $F-$term conditions of $\kappa^0$ and $\xi^{0}_{s}$ are
\begin{eqnarray}
\nonumber&&\frac{\partial w_{d}^{\text{sol}}}{\partial\kappa^{0}}=f_{1} (\chi_{s_{1}} \psi_{s_{1}}+\chi_{s_{2}} \psi_{s_{3}}+\chi_{s_{3}} \psi_{s_{2}})+ f_{2} (\varphi^2_{s_{1}} +2\varphi_{s_{2}} \varphi_{s_{3}})=0\,,\\
&&\frac{\partial w_{d}^{\text{sol}}}{\partial\xi^{0}_{s}}=M_{\xi_{s}}  \xi_{s}+f_{11} \left(\phi_{s_{1}}^2+2 \phi_{s_{2}} \phi_{s_{3}}\right)=0\,,
\end{eqnarray}
which together with Eq.~\eqref{eq:solar_VEV_1} lead to the following relations among the VEVs $v_{\xi_{s}}$, $v_{\eta_{s}}$, $v_{\chi_{s}}$, $v_{\psi_{s}}$, $v_{\varphi_{s}}$ and $v_{\phi_{s}}$
\begin{equation}
v^5_{\chi_{s}}=-\frac{f_{1}f_{5}f^3_{8}}{3f_{2}f_{6}f^3_{7}}v^5_{\eta_{s}}, \quad v_{\varphi_{s}}=\frac{f_{7}v^2_{\chi_{s}}}{f_{8}v_{\eta_{s}}},\quad v_{\psi_{s}}=\frac{f_{5}f_{8} v^3_{\eta_{s}}}{f_{6}f_{7}v^2_{\chi_{s}} }, \quad
v^2_{\phi_{s}}=-\frac{2 M_{\xi_{s}} v_{\xi_{s}} }{51 f_{11}}\,.
\end{equation}
Notice that the vacuum configurations of $\eta_{s}$, $\chi_{s}$, $\psi_{s}$, $\varphi_{s}$ and $\phi_{s}$ preserve the subgroup $Z^{U}_2$. In addition, the ratio $v^2_{\phi_{s}}/v_{\xi_{s}}=-\frac{2 M_{\xi_{s}} }{51 f_{11}}$ which contributes to the parameter $m_s$ is real because of the CP symmetry.

As regards the higher order corrections to the driving superpotential $w_d$, we note the operators comprising one driving field and three flavons are forbidden for the assignments in table~\ref{tab:bf}. The subleading contributions to $w^{l}_d$ which can shift the vacuum of $\eta_{l}$ and $\phi_{l}$ in Eq.~\eqref{eq:vacuum_ch} contain five flavons, and they are highly suppressed by $1/\Lambda^3$ with respect to the renormalizable terms. The subleading terms of $w^{\text{atm}}_d$ and $w^{\text{sol}}_d$ involve four flavon fields\footnote{A single operator $(\sigma^0\phi^3_{s})_{\mathbf{1}}$ is allowed by the symmetry of the model at order $\mathcal{O}(1/\Lambda)$. However, this term vanishes exactly when the vacuum alignment of $\phi_s$ in Eq.~\eqref{eq:vacuum_solar_3} is inserted.}, the resulting corrections are suppressed by $1/\Lambda^2$ compared to the contribution from the leading order terms and therefore can be safely neglected.

\subsection{\label{subsec:model}The structure of the model}

The lowest dimensional Yukawa operators invariant under the family symmetry $S_4\times Z_{5}\times Z_6 \times Z_8$,
responsible for the charged lepton masses,
are given by
\begin{eqnarray}
\nonumber  w_l&&=\frac{y_{\tau}}{\Lambda}\left(L\phi_{l}\right)_{\mathbf{1}}\tau^{c}H_d
+\frac{y_{\mu_1}}{\Lambda^2}\left(L\left(\eta_{l}\phi_{l}\right)_{\mathbf{3}}\right)_{\mathbf{1}}\mu^{c}H_d
+\frac{y_{\mu_2}}{\Lambda^2} \left(L\left(\phi_{l}\phi_{l}\right)_{\mathbf{3}}\right)_{\mathbf{1}}\mu^{c}H_d\\
\nonumber&&+\frac{y_{e_1}}{\Lambda^3}\left(L\phi_{l}\right)_{\mathbf{1}}\left(\eta_{l}\eta_{l}\right)_{\mathbf{1}}e^cH_d
+\frac{y_{e_2}}{\Lambda^3}\left(\left(L\phi_{l}\right)_{\mathbf{2}}\left(\eta_{l}\eta_{l}\right)_{\mathbf{2}}\right)_{\mathbf{1}}e^cH_d
+\frac{y_{e_3}}{\Lambda^3}\left(\left(L\eta_{l}\right)_{\mathbf{3}}\left(\phi_{l}\phi_{l}\right)_{\mathbf{3}}\right)_{\mathbf{1}}e^cH_d \\
\nonumber&&
+\frac{y_{e_4}}{\Lambda^3}\left(\left(L\eta_{l}\right)_{\mathbf{3}^{\prime}}\left(\phi_{l}\phi_{l}\right)_{\mathbf{3}^{\prime}}\right)_{\mathbf{1}}e^cH_d
 +\frac{y_{e_5}}{\Lambda^3}\left(L\phi_{l}\right)_{\mathbf{1}}\left(\phi_{l}\phi_{l}\right)_{\mathbf{1}}e^{c}H_d
+\frac{y_{e_6}}{\Lambda^3}\left(\left(L\phi_{l}\right)_{\mathbf{2}}\left(\phi_{l}\phi_{l}\right)_{\mathbf{2}}\right)_{\mathbf{1}}e^cH_d\\
\label{eq:ch_Yukawa}&&+\frac{y_{e_7}}{\Lambda^3}\left(\left(L\phi_{l}\right)_{\mathbf{3}}\left(\phi_{l}\phi_{l}\right)_{\mathbf{3}}\right)_{\mathbf{1}}e^cH_d
+\frac{y_{e8}}{\Lambda^3}\left(\left(L\phi_{l}\right)_{\mathbf{3}^{\prime}}\left(\phi_{l}\phi_{l}\right)_{\mathbf{3}^{\prime}}\right)_{\mathbf{1}}e^cH_d\,,
\end{eqnarray}
where all couplings are real since generalized CP symmetry is imposed on the model. Because the contraction $(\phi_{l}\phi_{l})_{\mathbf{3}}$ vanishes due to the antisymmetry of the associated Clebsch-Gordan coefficients, the terms proportional to $y_{\mu_2}$, $y_{e_3}$ and $y_{e_7}$ give null contributions. Plugging the VEVs of $\eta_{l}$ and $\phi_{l}$ in Eq.~\eqref{eq:vacuum_ch} into the above superpotential $w_{l}$, we find the charged lepton mass matrix is diagonal and the three charged lepton  masses are
\begin{eqnarray}
\nonumber&& m_e=\left|\left(y_{e_6}-2y_{e_8}-2y_{e_4}v_{\eta_{l}}/v_{\phi_{l}}+y_{e_2}v^2_{\eta_{l}}/v^2_{\phi_{l}}\right)\frac{v^3_{\phi_{l}}}{\Lambda^3}\right|v_d,\\
&&m_{\mu}=\left|y_{\mu_1}\frac{v_{\eta_{l}}v_{\phi_{l}}}{\Lambda^2}\right|v_d,\qquad m_{\tau}=\left|y_{\tau}\frac{v_{\phi_{l}}}{\Lambda}\right|v_d\,,
\end{eqnarray}
where $v_{d}=\langle H_{d}\rangle$. Note that auxiliary symmetry $Z_{5}\times Z_6 \times Z_8$ imposes different powers of $\eta_{l}$ and $\phi_{l}$ for the electron, muon and tau lepton mass terms. As a result, the electron, muon and tau masses arise at order $(\langle\Phi_{l}\rangle/\Lambda)^3$, $(\langle\Phi_{l}\rangle/\Lambda)^2$
and $\langle\Phi_{l}\rangle/\Lambda$ respectively, where $\Phi_{l}$ is either $\phi_{l}$ or $\eta_{l}$. The realistic mass hierarchy
can be reproduced if $\langle\Phi_{l}\rangle/\Lambda$ is of order $\lambda^2$, where $\lambda\simeq0.23$ denotes the Cabibbo angle. Moreover,
the subleading operators related to $e^{c}$, $\mu^{c}$ and $\tau^{c}$ comprise five flavons and consequently are suppressed by $1/\Lambda^5$. Such corrections have a minor impact on the results for the charged lepton masses and lepton mixing parameters and can be neglected.

In the neutrino sector, the leading order operators contributing to the neutrino masses are
\begin{equation}
\label{eq:w_nu}w_{\nu}=\frac{y_{a}}{\Lambda}\left(L\phi_{a}\right)_{\mathbf{1}}H_{u}\nu^{c}_{\text{atm}}+
\frac{y_{s}}{\Lambda}\left(L\phi_{s}\right)_{\mathbf{1^\prime}}H_{u}\nu^{c}_{\text{sol}}
+x_{a}\nu^{c}_{\text{atm}}\nu^{c}_{\text{atm}}\xi_{a}
+x_{s}\nu^{c}_{\text{sol}}\nu^{c}_{\text{sol}}\xi_{s}\,,
\end{equation}
where the coupling constants $y_{a}$, $y_{s}$, $x_{a}$ and $x_{s}$ are real parameters because of the imposed CP symmetry. The neutrino Dirac mass matrix $m_{D}$ arises from the first two terms in Eq.~\eqref{eq:w_nu}. With the vacuum alignments of $\phi_a$ and $\phi_s$ given in Eq.~\eqref{eq:vacuum_atm} and Eq.~\eqref{eq:vacuum_solar_3}, we find $m_{D}$ takes the following form
\begin{equation}\label{eq:MD_M1}
m_{D}=\begin{pmatrix}
y_{a}v_{\phi_{a}} ~&~ y_{s}v_{\phi_{s}} \\
\omega y_{a}v_{\phi_{a}}  ~&~ -\frac{7}{2}y_{s}v_{\phi_{s}} \\
\omega^2y_{a}v_{\phi_{a}}  ~&~ -\frac{7}{2}y_{s}v_{\phi_{s}}
\end{pmatrix}\frac{v_{u}}{\Lambda}\,,
\end{equation}
where $v_{u}=\langle H_{u}\rangle$. When the singlet flavons $\xi_{a}$ and $\xi_{s}$ obtain VEVs, the last two terms of $w_{\nu}$ lead to a diagonal right-handed neutrino mass matrix
\begin{equation}\label{eq:m_N}
m_{N}=\begin{pmatrix}
x_{a}v_{\xi_{a}}  &  0  \\
0  &  x_{s}v_{\xi_{s}}
\end{pmatrix}\,.
\end{equation}
Using the seesaw relation $m_{\nu}=-m_{D}m^{-1}_{N}m^{T}_{D}$, we can read off the light neutrino Majorana mass matrix
\begin{equation}
\label{eq:mnu_model}m_{\nu}=m_a\begin{pmatrix}
 1 &~ \omega  &~ \omega ^2 \\
 \omega  &~ \omega ^2 &~ 1 \\
 \omega ^2 &~ 1 &~ \omega  \\
\end{pmatrix}+m_se^{i\eta}\begin{pmatrix}
 1 &~  -7/2 &~  -7/2 \\
-7/2 &~ 49/4 &~ 49/4 \\
-7/2 &~ 49/4 &~ 49/4
\end{pmatrix}\,,
\end{equation}
with
\begin{equation}
m_a=-\frac{y^2_{a}v^2_{\phi_{a}}}{x_{a}v_{\xi_{a}}}\frac{v^2_{u}}{\Lambda^2},\qquad
m_se^{i\eta}=-\frac{y^2_{s}v^2_{\phi_{s}}}{x_{s}v_{\xi_{s}}}\frac{v^2_{u}}{\Lambda^2}\,.
\end{equation}
We see that the resulting neutrino mass matrix in Eq.~\eqref{eq:mnu_model} is of the same form as Eq.~\eqref{eq:mnu} but with fixed value $x=-7/2$. Moreover, both ratios $v^2_{\phi_{a}}/v_{\xi_{a}}$ and $v^2_{\phi_{s}}/v_{\xi_{s}}$ can be expressed in terms of the parameters of the driving superpotential and thus they are real, as shown in section~\ref{subsec:alignment}. As a consequence, the relative phase $\eta$ is either $0$ or $\pi$. The desired value $\eta=\pi$ can be achieved for $h_{1}f_{11}x_{a}x_{s}M_{\xi_{a}}M_{\xi_{s}}<0$.

Following the procedure outline in section~\ref{sec:new_LSS}, we find the lepton mixing matrix for $\eta=\pi$ takes the following form
\begin{equation}
U_{PMNS}=\frac{1}{5\sqrt{6}}\begin{pmatrix}
 7 \sqrt{2} ~&~ -2  \sqrt{13}\,i \cos \theta  ~&~ 2  \sqrt{13}  \sin \theta  \\
 \sqrt{26} ~&~ 7 i \cos \theta -5 \sqrt{3}  \sin \theta  ~&~ -7 \sin \theta +5\sqrt{3}\,i \cos \theta   \\
 \sqrt{26} ~&~ 7 i \cos \theta +5\sqrt{3}  \sin \theta   ~&~ -7\sin \theta -5 \sqrt{3}\,i \cos \theta
\end{pmatrix}\,,
\end{equation}
with
\begin{equation}
\sin2\theta=\frac{10 |14r-1|}{13 \sqrt{289 r^2+32 r+4}}, \quad \cos2\theta=\frac{3 \left(57r+8\right)}{13 \sqrt{ 289 r^2+32 r+4}}\,,
\end{equation}
where $r=m_s/m_a$. Then the analytical expressions for the lepton mixing angles can be extracted,
\begin{equation}
\label{eq:mixing_angles_benchmark}\sin^2\theta_{13}=\frac{26  }{75}\sin ^2\theta,\quad
\sin^2\theta_{12}=\frac{26 \cos ^2\theta }{62+13 \cos 2 \theta },\quad
\sin^2\theta_{23}=\frac{1}{2}\,.
\end{equation}
We see that the atmospheric angle $\theta_{23}$ is maximal, the solar and the reactor mixing angles fulfill the sum rule
\begin{equation}
\label{eq:sum_rule_benchmark}\cos^2\theta_{12}\cos^2\theta_{13}=\frac{49}{75}\,.
\end{equation}
Using the $3\sigma$ interval $0.01981\leq\sin^2\theta_{13}\leq0.02436$~\cite{Esteban:2016qun}, from Eq.~\eqref{eq:mixing_angles_benchmark} we find the allowed range of the parameter $r$ is $0.448\leq r\leq0.634$ and the exact sum rule in Eq.~\eqref{eq:sum_rule_benchmark} gives $0.330\leq\sin^2\theta_{12}\leq0.333$ which can be tested by JUNO in near future~\cite{An:2015jdp}. For the CP invariants, we get
\begin{equation}
J_{CP}=-\frac{91}{750\sqrt{3}} \sin2\theta,\quad I_1=0\,.
\end{equation}
Therefore the Dirac phase $\delta_{CP}$ is maximal and the Majorana phase $\beta$ is trivial with
\begin{equation}
\delta_{CP}=-0.5\pi,\quad \beta=0\,.
\end{equation}
Furthermore, we report the exact results for the neutrino masses
\begin{eqnarray}
\nonumber && m^2_2=\frac{9}{8} m^2_{a}\left(289 r^2-18 r+4-\left| 2-17 r\right|\sqrt{289 r^2+32 r+4}  \right)\,, \\
&& m^2_3=\frac{9}{8} m^2_{a}\left(289 r^2-18 r+4+\left| 2-17 r\right|\sqrt{289 r^2+32 r+4}  \right)\,,
\end{eqnarray}
with the lightest neutrino massless $m_1=0$. For the best fit values of $m_{a}=3.720\,\text{meV}$ and $r=0.553$, the neutrino masses $m_{2}$ and $m_{3}$ are
\begin{equation}
 m_{2}= 8.622 \,\text{meV}, \qquad m_{3}= 49.918 \,\text{meV}\,,
\end{equation}
as given in table~\ref{tab:bf}. We note that the next-to-leading operators of $w_{\nu}$ are $\nu^{c}_{\text{atm}}\nu^{c}_{\text{atm}}(\phi^2_{a})_{\mathbf{1}}$, $\nu^{c}_{\text{sol}}\nu^{c}_{\text{sol}}(\phi^2_{s})_{\mathbf{1}}$ and $\nu^{c}_{\text{atm}}\nu^{c}_{\text{sol}}(\phi_{a}\phi_{s})_{\mathbf{1^\prime}}$. The contributions of the first two terms can be absorbed via a redefinition of the parameters $x_{a}$ and $x_s$ since both $v^2_{\phi_{a}}/v_{\xi_{a}}$ and $v^2_{\phi_{s}}/v_{\xi_{s}}$ are real. The last term will generate off-diagonal elements of the right-handed neutrino mass matrix. The corresponding corrections to the leading order results for the mixing parameters are of relative order $\lambda^2$. In summary, we have reproduced the benchmark model of the new LSS variants highlighted in table~\ref{tab:bf}.

\section{\label{sec:Conclusion} Conclusion}

In this paper we have proposed a new {\it tri-direct CP approach} for two right-handed neutrino models based on the idea that the high energy family and CP symmetry $G_f\rtimes H_{CP}$ is spontaneously broken down to $G_{\text{atm}}\rtimes H_{CP}^{\text{atm}}$ in the sector of one of the right-handed neutrinos, and $G_{\text{sol}}\rtimes H_{CP}^{\text{sol}}$ in the sector of the other right-handed neutrino, with the charged lepton sector having a residual flavour symmetry $G_{l}$, as illustrated in figure~\ref{fig:benz}.

In such a {\it tri-direct CP approach} we have shown that the combination of the three residual symmetries provides a new way of fixing the parameters. In particular it can lead to vacuum alignments in the neutrino sector which are uniquely fixed by symmetry, unlike the {\it semi-direct CP approach} where not all such vacuum alignments are uniquely fixed. To illustrate the approach, we have revisited the Littlest Seesaw model based on $S_4$ and shown that the {\it tri-direct CP approach} based on $G_{\text{atm}}=Z^{U}_2$ and $G_{\text{sol}}=Z^{SU}_2$
uniquely fixes alignments which are not uniquely fixed in the {\it semi-direct CP approach} based on a common residual symmetry $G_{\nu}=Z^{SU}_2$ in the neutrino sector.

Following the {\it tri-direct CP approach}, we have also proposed new variants of the Littlest Seesaw model which have not so far appeared in the literature, with different predictions for each variant. We have performed a comprehensive numerical analysis of a selection of benchmark points within the LSS variants arising from $S_4$, in order to determine their viability and predictions. Although the benchmarks within most of the variants have a larger $\chi^2$ than the original LSS model, which provides an excellent agreement with experimental data, one of the benchmarks has a relatively low $\chi^2\approx 4.5$. We have proposed an explicit model which can realise this successful benchmark point, based on the atmospheric
flavon vacuum alignment $(1, \omega^2 , \omega)$ and the solar flavon vacuum alignment $(1, -7/2, -7/2 )$. The model has exact accidental $\mu \tau $ reflection symmetry~\cite{King:2018kka} and hence predicts maximal atmospheric mixing and maximal Dirac CP violation.

Although the flavor symmetry and CP symmetry are completely broken in the whole neutrino sector, the tri-direct CP models are rather predictive. The light neutrino mass matrix is generally predicted to depend on only four parameters $m_a$, $m_s$, $x$ and $\eta$, and the last two parameters $x$ and $\eta$ can be fixed to certain benchmark values by explicit superpotential alignment terms in a concrete model. The cross product $\langle\phi_{\text{atm}}\rangle\times\langle\phi_{\text{sol}}\rangle$ is an eigenvector of $m_{\nu}$ with zero eigenvalue, and neutrino mass spectrum can be either normal ordering or inverted ordering. Finally we note that the tri-direct CP approach can be extended to the case of three right-handed neutrinos, then the flavons associated with each right-handed neutrino in the Yukawa couplings preserve different residual symmetries. Such a scenario can also lead to phenomenologically viable lepton mixing angles and neutrino masses.

\subsection*{Acknowledgements}
G.-J.\, D. acknowledges the support of the National Natural Science Foundation of China under Grant No 11522546.
S.\,F.\,K. acknowledges the STFC Consolidated Grant ST/L000296/1
and the European Union's Horizon 2020 research and innovation programme under the Marie Sk\l{}odowska-Curie grant agreements
Elusives ITN No.\ 674896 and InvisiblesPlus RISE No.\ 690575. C.-C.\, L. is supported by China Postdoctoral Science Foundation  Grant Nos. 2017M620258 and 2018T110617, CPSF-CAS Joint Foundation for Excellent Postdoctoral Fellows No. 2017LH0003,   the Fundamental Research Funds for the Central Universities under Grant No. WK2030040090 and the CAS Center for Excellence in Particle Physics (CCEPP).

\newpage

\section*{\label{sec:appendix}Appendix}

\begin{appendix}

\section{\label{sec:appendix_A}Diagonalization of the neutrino mass matrix $m^\prime_{\nu}$ }

In this appendix, we will present the results for the diagonalization of  $m^\prime_{\nu}$. From Eqs.~\eqref{eq:nu_mass_matrix2} and \eqref{eq:yzw_par}, we find the neutrino mass matrix $m^\prime_{\nu}$ can be written as
\begin{equation}
m^\prime_{\nu}=
\begin{pmatrix}
0 & ~ 0 &~ 0 \\
0 &~ |y|e^{i\phi_{y}}  ~&~  |z|e^{i\phi_{z}} \\
0 &~ |z|e^{i\phi_{z}}  ~&~  |w|e^{i\phi_{w}}
\end{pmatrix}\,,
\end{equation}
where the parameters $y$, $z$, $w$, $\phi_{y}$, $\phi_{z}$ and $\phi_{w}$ are given in Eq.~\eqref{eq:yzw_par}. The neutrino mass matrix $m'_{\nu}$  can be exactly diagonalized by a unitary matrix $U_{\nu2}$~\cite{Ding:2013bpa},
\begin{equation}
\label{eq:diag_2nd}U^{T}_{\nu2}m^{\prime}_{\nu}U_{\nu2}=\text{diag}(0, m_2, m_3) \,.
\end{equation}
The light neutrino masses $m_2$ and $m_3$ are given by
\begin{equation}\label{eq:nu_masses}
m^2_2=\frac{1}{2}\left[|y|^2+|w|^2+2|z|^2-\frac{|w|^2-|y|^2}{\cos2\theta}\right], \quad
m^2_3=\frac{1}{2}\left[|y|^2+|w|^2+2|z|^2+\frac{|w|^2-|y|^2}{\cos2\theta}\right]\,,
\end{equation}
where the rotation angle $\theta$ is specified by
\begin{eqnarray}
\nonumber&&\sin2\theta=\frac{2|z|\sqrt{|y|^2+|w|^2+2|y||w|\cos(\phi_{y}+\phi_{w}-2\phi_{z})}}
{\sqrt{(|w|^2-|y|^2)^2+4|z|^2\left[|y|^2+|w|^2+2|y||w|\cos(\phi_{y}+\phi_{w}-2\phi_{z})\right]}},\\
\label{eq:theta}&&\cos2\theta=\frac{|w|^2-|y|^2}{\sqrt{(|w|^2-|y|^2)^2+4|z|^2
\left[|y|^2+|w|^2+2|y||w|\cos(\phi_{y}+\phi_{w}-2\phi_{z})\right]}}\,.
\end{eqnarray}
The unitary matrix $U_{\nu2}$ in Eq.~\eqref{eq:diag_2nd} takes the following form,
\begin{equation}\label{eq:Unu2}
U_{\nu2}=
\begin{pmatrix}
1  &~    0    &~   0 \\
0  &~  \cos\theta \,e^{i(\psi+\rho)/2}   &~  \sin\theta \,e^{i(\psi+\sigma)/2}   \\
0  &~   -\sin\theta\, e^{i(-\psi+\rho)/2}    &~~    \cos\theta\, e^{i(-\psi+\sigma)/2}
\end{pmatrix} \,,
\end{equation}
where the phases $\psi$, $\rho$ and $\sigma$ are expressed in terms of model parameters as
\begin{small}
\begin{eqnarray}
\nonumber&&\sin\psi=\frac{-|y|\sin(\phi_{y}-\phi_{z})+|w|\sin(\phi_{w}-\phi_{z})}
{\sqrt{|y|^2+|w|^2+2|y||w|\cos(\phi_{y}+\phi_{w}-2\phi_{z})}}\,, \\
\nonumber && \cos\psi=\frac{|y|\cos(\phi_{y}-\phi_{z})+|w|\cos(\phi_{w}-\phi_{z})}{\sqrt{|y|^2+|w|^2+2|y||w|\cos(\phi_{y}+\phi_{w}-2\phi_{z})}}\,,\\
\nonumber&&\sin\rho=-\frac{(m^2_2-|z|^2)\sin\phi_{z}+|y||w|\sin(\phi_{y}+\phi_{w}-\phi_{z})}
{m_2\sqrt{|y|^2+|w|^2+2|y||w|\cos(\phi_{y}+\phi_{w}-2\phi_{z})}}\,,\\
\nonumber && \cos\rho=\frac{(m^2_2-|z|^2)\cos\phi_{z}+|y||w|\cos(\phi_{y}+\phi_{w}-\phi_{z})}
{m_2\sqrt{|y|^2+|w|^2+2|y||w|\cos(\phi_{y}+\phi_{w}-2\phi_{z})}}\,,\\
\nonumber &&\sin\sigma=-\frac{(m^2_3-|z|^2)\sin\phi_{z}+|y||w|\sin(\phi_{y}+\phi_{w}-\phi_{z})}
{m_3\sqrt{|y|^2+|w|^2+2|y||w|\cos(\phi_{y}+\phi_{w}-2\phi_{z})}}\,,\\
\label{eq:prs_UGR1_Td} && \cos\sigma=\frac{(m^2_3-|z|^2)\cos\phi_{z}+|y||w|\cos(\phi_{y}+\phi_{w}-\phi_{z})}
{m_3\sqrt{|y|^2+|w|^2+2|y||w|\cos(\phi_{y}+\phi_{w}-2\phi_{z})}}\,.
\end{eqnarray}
\end{small}

\end{appendix}

\providecommand{\href}[2]{#2}\begingroup\raggedright\endgroup

\end{document}